\documentclass[12pt]{article}

\usepackage{amsmath}
\usepackage{graphicx}
\usepackage{subfig}

\usepackage{color}

\newcommand {\apgt} {\ {\raise-.5ex\hbox{$\buildrel>\over\sim$}}\ }

\begin{document}

\begin{titlepage}

\vspace{-4cm}

\title{
   {\LARGE The $B$-$L$/Electroweak Hierarchy \\[.1in]
    in Smooth Heterotic Compactifications\\[.5cm]  }}
                       
\author{{\bf
   Michael Ambroso$^{1}$ 
   and Burt A.~Ovrut$^{1,2}$}\\[5mm]
   {\it $^1$Department of Physics, University of Pennsylvania} \\
   {\it Philadelphia, PA 19104--6396}\\[2mm]
   {\it $^2$School of Natural Sciences, The Institute for Advanced Study} \\
   {\it Princeton, New Jersey, 08540}} 
\date{}

\maketitle

\begin{abstract}
\noindent

$E_{8} \times E_{8}$ heterotic string and M-theory, when appropriately compactified, can give rise to realistic, $N=1$ supersymmetric  particle physics. In particular, the exact matter spectrum of the MSSM, including three right-handed neutrino supermultiplets, one per family, and one pair of Higgs-Higgs conjugate superfields is obtained by compactifying on Calabi-Yau manifolds admitting specific $SU(4)$ vector bundles. These ``heterotic standard models'' 
have the $SU(3)_{C} \times SU(2)_{L} \times U(1)_{Y}$ gauge group of the standard model augmented by an additional gauged $U(1)_{B-L}$. Their minimal content requires that the $B$-$L$ gauge symmetry be spontaneously broken by a vacuum expectation value of at least one right-handed sneutrino. In a previous paper, we presented the results of a renormalization group analysis showing that $B$-$L$ gauge symmetry is indeed radiatively broken with a
$B$-$L$/electroweak hierarchy of ${\cal{O}}(10)$ to ${\cal{O}}(10^{2})$.  In this paper, we present the details of that analysis, extending the results to include higher order terms in $tan\beta^{-1}$ and the explicit spectrum of all squarks and sleptons.

\vspace{.3in}
\noindent
\end{abstract}

\thispagestyle{empty}

\end{titlepage}

\section{Introduction}

Smooth compactifications of the weakly coupled \cite{Aa} and strongly coupled \cite{Ba,Bb,Bc} $E_{8} \times E_{8}$ heterotic string have been studied for many years. When the compactification is on a Calabi-Yau threefold with a slope-stable, holomorphic vector bundle, the low energy four-dimensional effective theory is $N=1$ supersymmetric. In recent years, such compactifications have been extended to complete intersection and elliptically fibered Calabi-Yau spaces admitting vector bundles constructed using monads \cite{Ca,Cb,Cc,Cd,Ce}, spectral covers \cite{Gaaa,Gaa,Ga,Gc} and by extension of lower rank bundles \cite{Hb,Hd}. The formalism for explicitly computing the low energy spectrum in each of these cases has been developed, and presented in \cite{Ia,Ib}, \cite{Ja,Jb} and \cite{Kb,Kc} respectively. Cohomological methods have been used to calculate the texture of Yukawa couplings and other parameters in these contexts  \cite{Lb,Lc,Ld}. Finally, the non-perturbative string instanton contributions to the superpotential have been computed \cite{Maaa,Maa,Ma,Mc} and used to discuss moduli stability, supersymmetry breaking and the cosmological constant \cite{Nb}. These methods underlie the theory of ``brane universes'' \cite{Oa,Ob} and new approaches to cosmology \cite{Pa,Pc}.

In a series of papers, compactifications of the $E_{8} \times E_{8}$ superstring have been constructed on elliptically fibered Calabi-Yau spaces with ${\bf Z}_{3} \times {\bf Z}_{3}$ homotopy over a 
$dP_{9}$ surface \cite{Qa}. These spaces admit a specific class of slope-stable, holomorphic vector bundles with structure group $SU(4)$ that are constructed by extension and are equivariant under ${\bf Z}_{3} \times {\bf Z}_{3}$ \cite{Ra}. The non-trivial homotopy allows one to extend these bundles with flat ${\bf Z}_{3} \times {\bf Z}_{3}$ Wilson lines. Using the methods referenced above, the four-dimensional spectrum of these theories was computed. It is found to have precisely the matter content of the minimal supersymmetric standard model (MSSM), including three right-handed neutrino chiral supermultiplets, one per family. In addition there are a small number of Higgs-Higgs conjugate pairs, two in the model presented in \cite{Sa} and one in the vacuum of \cite{Ta}. These are termed ``heterotic standard models''. They all contain a relatively small number of geometric and vector bundle moduli and each possesses an acceptable texture of Yukawa couplings. For specificity, in this paper we will consider the minimal heterotic standard model \cite{Ta} containing the exact matter spectrum of the MSSM, one pair of Higgs-Higgs conjugate superfields as well as three complex structure moduli, three Kahler moduli and thirteen vector bundle moduli.

The four-dimensional gauge group is obtained through the sequential breaking of $E_{8}$ by the 
$SU(4)$ structure group of the vector bundle and the ${\bf Z}_{3} \times {\bf Z}_{3}$ of the Wilson lines. We find that
\begin{equation}
E_{8} \stackrel{SU(4)}{\longrightarrow} Spin(10)  \stackrel{Z_{3} \times Z_{3}}{\longrightarrow} 
SU(3)_{C} \times SU(2)_{L} \times U(1)_{Y} \times U(1)_{B-L} \ .
\nonumber
\end{equation}
Note that in addition to the standard model gauge group $SU(3)_{C} \times SU(2)_{L} \times U(1)_{Y}$, there is an extra gauged $U(1)_{B-L}$. This arises from the fact that $Spin(10)$ has rank five and that the rank must be preserved when the group is further broken by any Abelian finite group, such as 
${\bf Z}_{3} \times {\bf Z}_{3}$. Since the standard model group has rank four, an extra $U(1)$ gauge factor must appear, in this case precisely $U(1)_{B-L}$. 

We would like to emphasize that there is a direct relationship between having three right-handed neutrino chiral multiplets, one per family, and the appearance of the additional $U(1)_{B-L}$ gauge factor. The $SU(4)$ structure group in heterotic standard models is chosen precisely because the decomposition of the $\bf 248 \rm$ of $E_{8}$ with respect to it contains the $\bf 16 \rm$ representation of $Spin(10)$. It is well-known that each $\bf 16 \rm$ is composed of one family of quarks and leptons, including a right-handed neutrino. This fact makes choosing a vector bundle with an $SU(4)$ structure group a natural way to ensure that the spectrum contains three right-handed neutrinos. However, the rank of $Spin(10)$ is five, one larger than the standard model gauge group. Hence, when $Spin(10)$ is broken to $SU(3)_{C} \times SU(2)_{L} \times U(1)_{Y}$ by the Abelian ${\bf Z}_{3} \times {\bf Z}_{3}$ Wilson lines, an additional $U(1)_{B-L}$ must appear.

Interestingly, the appearance of this $U(1)_{B-L}$ factor naturally solves another important phenomenological problem. As is well-known, the MSSM must be augmented by a specific ${\bf Z}_{2}$ symmetry called $R$-parity (or matter parity, its supersymmetric extension) \cite{Ud,Ua,Ub,Uc}. This prohibits dangerous baryon and lepton number violating operators that would lead, for example, to rapid nucleon decay. Unfortunately, it has proven difficult to obtain discrete matter parity as a symmetry of smooth heterotic compactifications. As we have seen, however, a gauged $U(1)_{B-L}$ symmetry arises naturally in compactifications with $SU(4)$ structure group. $U(1)_{B-L}$ contains matter parity as a subgroup and, hence, as long as $U(1)_{B-L}$ is unbroken, matter parity is a symmetry. Since $U(1)_{B-L}$ is not observed in electroweak processes,  it must be spontaneously broken at a higher scale. In supersymmetric grand unified theories (GUTs), where the matter content can be chosen arbitrarily, this breaking can occur at a high scale \cite{Va,Vb,Vc,Vd,Ve}. This is accomplished by adding multiplets to the MSSM, neutral under the standard model group, for which $3(B-L)$ is an even, non-zero integer \cite{Vf}.  It is then arranged for at least one of these multiplets to get a large vacuum expectation value (VEV). This spontaneously breaks the gauged symmetry, giving a large mass to the $B$-$L$ vector boson, but, since the $3(B-L)$ charge is even, leaves ${\bf Z}_{2}$ matter parity as a discrete symmetry. Hence, although $U(1)_{B-L}$ is broken at a high scale, ${\bf Z}_{2}$ matter parity is an exact symmetry at lower scales. However, this mechanism cannot occur within the context of the exact MSSM spectrum in the heterotic standard models, since for all fields $3(B-L)$ is either $\pm 1$, $\pm3$ or $0$. It follows that when $U(1)_{B-L}$ is broken by a non-zero VEV, there is no residual ${\bf Z}_{2}$ symmetry. Therefore, in heterotic standard models $U(1)_{B-L}$ gauge symmetry must be broken at a low scale, an order of magnitude or two above the electroweak scale. But how can this be accomplished? 

In the MSSM spectrum, the only scalar fields that carry a non-trivial $B$-$L$ charge, but transform trivally under the standard model gauge group, are the right-handed sneutrinos. Thus, $B$-$L$ symmetry must be broken by at least one of these scalars acquiring a non-vanishing VEV from radiative corrections. However, to sufficiently suppress large baryon and lepton number violation, this must  occur just above the electroweak scale. To analyze this, one adds to the supersymmetric MSSM the soft supersymmetry violating operators that arise from various sources, such as gaugino condensation and the moduli vacuum state, during compactification. Whatever the source, these operators are of a specific form first worked out in \cite{W} and discussed within the context of generalized scenarios in \cite{Xa,Xb,Xc,Xd,Xe}. The initial values of the parameters are set by the details of the compactification, and are generically moduli dependent. At any lower scale, these parameters are determined by a complicated set of intertwined, non-linear renormalization group equations (RGEs) \cite{Ya,Yb,Yc,Yd,Ye,Yf,Yg}. It is by no means clear that a non-vanishing sneutrino VEV will necessarily develop. If it does, one must still show that a neutral Higgs field will get a non-zero VEV, thus breaking electroweak symmetry, at a scale an order of magnitude or two lower than the sneutrino VEV. Finally, it is of interest to know whether this result requires extremely fine-tuned parameters or is, more or less, a natural hierarchy.

In a previous paper \cite{Z}, we presented the results of a renormalization group analysis of the minimal heterotic standard model with a reasonable set of assumptions about the initial soft supersymmetry breaking parameters. These assumptions are consistent with basic requirements of phenomenology, such as suppressed flavor changing neutral currents, but are further constrained so as to allow a quasi-analytic solution of the RGEs. We found that $B$-$L$ symmetry is indeed spontaneously broken by a radiatively induced VEV of at least one right-handed sneutrino. Electroweak symmetry is then radiatively broken by a Higgs VEV at a lower scale, with the $B$-$L$/electroweak hierarchy of ${\cal{O}}(10)$ to ${\cal{O}}(10^{2})$. The purpose of this paper is to present the detailed renormalization group calculations leading to those conclusions. These include both analytic, quasi-analytic and purely numerical solutions of the relevant equations. 

Specifically, we will do the following. In Section 2, the chiral fields of the $U(1)_{B-L}$ extended MSSM are presented and their supersymmetric interactions, via the superpotential and $D$-terms, are discussed. The complete set of soft supersymmetry breaking operators in this context are then introduced. Section 3 is devoted to presenting and solving the RGEs associated with the spontaneous breaking of the $U(1)_{B-L}$ gauge symmetry. This analysis involves the gauge parameters,  gaugino and slepton masses and both the $B$-$L$ and $Y$ Fayet-Iliopoulos (FI) parameters \cite{FI}. Using these results, it is shown that $U(1)_{B-L}$ is indeed radiatively broken by a non-zero right-handed sneutrino VEV, and the details of this vacuum are presented. In Section 4, the analysis is extended to include both up and down Higgs and squark masses, as well as the $\mu$ and $B$ parameters. It is then shown that at the $B$-$L$ scale, electroweak symmetry, as well as color and charge, remain unbroken. All RGEs are then scaled down several orders of magnitude. We demonstrate that a Higgs VEV now develops which spontaneously breaks electroweak symmetry, without breaking color or charge. The $B$-$L$ and Higgs VEVs are presented and all squark, slepton and Higgs masses are calculated in this vacuum. The results are a detailed function of the initial right-handed sneutrino  and Higgs mass parameters, $m_{\nu}(0)$ and $m_{H}(0)$ respectively, as well as inverse powers of $tan\beta$. The relationship between $m_{\nu}(0)$ and $m_{H}(0)$ is also presented. Finally, in Section 5, we analyze the resultant $B$-$L$/electroweak hierarchy and show that it is of order $10$ to $10^{2}$. The complete spectrum of squark and slepton masses, written in terms of the $B$-$L$ boson mass, is then evaluated. Our analysis depends on a numerical solution for the Higgs mass parameter, $m_{H}(t)^{2}$. This is discussed and presented in Appendix A. In Appendix B, we numerically calculate the relationship between $m_{\nu}(0)$ and $m_{H}(0)$ used in the text. Finally, in Appendix C we verify that our $B$-$L$/electroweak breaking vacuum satisfies the standard constraint and minimization equations presented, for example, in~\cite{Yg,Zb,Zc}. 

\section{The $N=1$ Supersymmetric Theory}

We will consider an $N=1$ supersymmetric theory with gauge group
\begin{equation}
G=SU(3)_{C} \times SU(2)_{L} \times U(1)_{Y} \times U(1)_{B-L}
\label{1}
\end{equation}
and the associated vector superfields. The gauge parameters are denoted by $g_{3}$, $g_{2}$, $g_{Y}$ and $g_{B-L}$ respectively. The matter spectrum consists of three families of quark and lepton chiral superfields, each family with a {\it right-handed neutrino}. They transform under the gauge group in the standard manner as
\begin{equation}
Q_{i}=({\bf 3},{\bf 2},1/3,1/3), \quad u_{i}=({\bf \bar{3}}, {\bf 1}, -4/3, -1/3), \quad d_{i}=({\bf \bar{3}}, {\bf 1}, 2/3, -1/3)
\label{2}
\end{equation}
for the left and right-handed quarks and
\begin{equation}
L_{i}=({\bf 1},{\bf 2},-1,-1), \quad \nu_{i}=({\bf 1}, {\bf 1}, 0, 1), \quad e_{i}=({\bf 1}, {\bf 1}, 2, 1)
\label{3}
\end{equation}
for the left  and right-handed leptons, where $i=1,2,3$. In addition, the spectrum has one pair of Higgs-Higgs conjugate chiral superfields transforming as
\begin{equation}
H=({\bf 1},{\bf 2},1,0), \qquad \bar{H}=({\bf 1},{\bf 2}, -1,0).
\label{4}
\end{equation}
When necessary, the left-handed $SU(2)_{L}$ doublets will be written as 
\begin{equation}
Q_{i}=(U_{i}, D_{i}), \quad L_{i}=(N_{i}, E_{i}), \quad H=(H^{+},H^{0}), \quad \bar{H}=({\bar{H}}^{0}, {\bar{H}}^{-}).
\label{5}
\end{equation}
There are {\it no other fields in the spectrum}.

The supersymmetric potential energy is given by the usual sum over the modulus squared of the $F$
and $D$-terms. In principle, the $F$-terms are determined from the most general superpotential  invariant under the gauge group,
\begin{equation}
W=\mu H\bar{H} +{\sum_{i,j=1}^{3}}\left(\lambda_{u, ij} Q_{i}Hu_{j}+\lambda_{d, ij} Q_{i}\bar{H}d_{j}+\lambda_{\nu, ij} L_{i}H\nu_{j}+\lambda_{e, ij} L_{i}\bar{H}e_{j}\right)
\label{6}
\end{equation}
Note that an innocuous mixing term of the form $L_{i}H$,
as well as the dangerous lepton and baryon number violating interactions
\begin{equation}
L_{i}L_{j}e_{k}, \quad L_{i}Q_{j}d_{k}, \quad u_{i}d_{j}d_{k}
\label{7}
\end{equation}
which generically would lead, for example, to rapid nucleon decay, are disallowed by the $U(1)_{B-L}$ gauge symmetry. To simplify the upcoming calculations, we will assume that we are in a mass-diagonal basis where
\begin{equation}
\lambda_{u, ij}=\lambda_{d, ij}=\lambda_{\nu, ij}=\lambda_{e, ij}=0, \quad i \neq j.
\label{8}
\end{equation}
Note that once these off-diagonal couplings vanish just below the compactification scale, they will do so at all lower energy-momenta. We will denote the diagonal Yukawa couplings by $\lambda_{ii}=\lambda_{i}$, $i=1,2,3$. 

Next, observe that a constant, field-independent $\mu$ parameter cannot arise in a supersymmetric string vacuum  since the Higgs fields are zero modes. However, the $H{\bar{H}}$ bilinear can have higher-dimensional couplings to moduli through both holomorphic and non-holomorphic interactions in the superpotential and Kahler potential respectively. When moduli acquire VEVs due to non-perturbative effects, these can induce non-vanishing supersymmetric contributions to $\mu$. A non-zero $\mu$ can also be generated by gaugino condensation in the hidden sector. Why this induced $\mu$-term should be small enough to be consistent with electroweak symmetry breaking is a difficult, model dependent problem. In this paper, we will not discuss this ``$\mu$-problem'', but simply assume that the $\mu$ parameter is at, or below, the electroweak scale. In fact, so as {\it to emphasize the $B$-$L$/electroweak hierarchy and simplify the calculation, we will take $\mu$, while non-zero, to be substantially smaller than the electroweak scale}, making its effect  sub-dominant. This can be implemented consistently throughout the entire scaling regime. The exact meaning of ``sub-dominant'' is quantified in Appendix C, where we also present the upper bound on $\mu$ and, hence, the Higgsino mass in our approximation scheme.

The $SU(3)_{C}$ and $SU(2)_{L}$ $D$-terms are of the standard form. We present the $U(1)_{Y}$ 
and $U(1)_{B-L}$ $D$-terms,
\begin{equation}
D_{Y}= \xi_{Y} +g_{Y}{\phi}_{A}^{\dagger}\left({\bf{Y}\rm}/2\right)_{AB}{\phi}_{B}
\label{9}
\end{equation}
and 
\begin{equation}
D_{B-L}= \xi_{B-L} +g_{B-L}{\phi}_{A}^{\dagger}\left({\bf{Y_{B-L}}\rm}\right)_{AB}{\phi}_{B}
\label{10}
\end{equation}
where the index $A$ runs over all scalar fields ${\phi}_{A}$, to set the notation for the hypercharge and $B$-$L$ charge generators and to remind the reader that each of these $D$-terms potentially has a Fayet-Iliopoulos additive constant.  However, as with the $\mu$ parameter, constant field-independent FI terms cannot occur in string vacua since the low energy fields are zero modes. Field-dependent FI terms can occur in some contexts, see for example \cite{Lara}. However, since both the hypercharge and $B$-$L$ gauge symmetries are anomaly free, such field-dependent FI  terms are not generated in the supersymmetric effective theory. We include them in (\ref{9}),(\ref{10}) since they can, in principle, arise at a lower scale from radiative corrections once supersymmetry is softly broken \cite{Yd}. Be that as it may, if calculations are done {\it in the $D$-eliminated formalism, which we use in this paper, these FI parameters can be consistently absorbed into the definition of the soft scalar masses} and their beta functions. Hence, we will no longer consider them.

In addition to the supersymmetric potential, the Lagrangian density also contains explicit ``soft'' supersymmetry violating terms. These arise from the spontaneous breaking of supersymmetry in  
a hidden sector that has been integrated out of the theory. This breaking can occur in either $F$-terms, $D$-terms or both in the hidden sector. In this paper, for simplicity, we will restrict our discussion to soft supersymmetry breaking terms arising exclusively from  $F$-terms. The form of these terms is well-known and, in the present context, given by \cite{W,Xa,Xb,Xd,Zc}
\begin{equation}
V_{\rm soft}=V_{2s}+V_{3s}+V_{2f},
\label{11}
\end{equation}
where $V_{2s}$ are scalar mass terms 
\begin{eqnarray}
V_{2s}  =& { \sum_{i=1}^{3}}& (m^{2}_{Q_{i}}|{Q}_{i}|^{2}+m^{2}_{u_{i}}|{u}_{i}|^{2}+
                        m^{2}_{d_{i}}|{d}_{i}|^{2} \nonumber \\ 
              &    & +m^{2}_{L_{i}}|{L}_{i}|^{2}+m^{2}_{\nu_{i}}|{\nu}_{i}|^{2}
                        +m^{2}_{e_{i}}|{e}_{i}|^{2} \label{12} \\   
              &    &+m_{H}^{2}|H|^{2} +m_{\bar{H}}^{2}|\bar{H}|^{2})  -(BH\bar{H}+hc), \nonumber   
\end{eqnarray}
$V_{3s}$ are scalar cubic couplings
\begin{equation}
V_{3s}=\sum_{i=1}^{3} (A_{u_{i}} {Q}_{i}H{u}_{i} +A_{d_{i}} {Q}_{i}{\bar{H}}{b}_{i} +A_{\nu_{i}} {L}_{i}H{\tilde{\nu}}_{i}+A_{e_{i}} {L}_{i}{\bar{H}}{e}_{i} +{\rm hc})
\label{13}
\end{equation}
and $V_{2f}$ contains the gaugino mass terms
\begin{equation}
V_{2f}= \frac{1}{2} M_{3} \lambda_{3} \lambda_{3}+ \frac{1}{2} M_{2} \lambda_{2} \lambda_{2}+ \frac{1}{2} M_{Y} \lambda_{Y} \lambda_{Y}+ \frac{1}{2} M_{B-L} \lambda_{B-L} \lambda_{B-L}+
{\rm hc}.
\label{14}
\end{equation}
As above, to simplify the calculation we assume the parameters in (\ref{12}) and (\ref{13}) are flavor-diagonal. This is consistent since once the off-diagonal parameters vanish just below the compactification scale, they will do so at all lower energy-momenta.

\section{The Renormalization Group and $B$-$L$}

In this section, we discuss the spontaneous breakdown of the gauged $B$-$L$ symmetry.
The parameters in our theory all scale with energy-momentum, each obeying the associated renormalization group equation (RGE). In this section, we will solve those equations required in the analysis of $B$-$L$ breaking to {\it the one-loop level}. 

\subsection*{Gauge Parameters:}

We begin by considering the RG running of the gauge coupling parameters. Since our low energy theory arises from an $SO(10)$ compactification of heterotic string theory broken to $SU(3)_{C} \times SU(2)_{L} \times U(1)_{Y} \times U(1)_{B-L}$ by Wilson lines, it is conventional to redefine the hypercharge and $B$-$L$ gauge parameters as 
\begin{equation}
g_{1}=\sqrt{\frac{5}{3}}g_{Y},  \qquad g_{4}=\sqrt{\frac{4}{3}}g_{B-L} \ .
\label{15}
\end{equation}
With these redefinitions, and defining $t=ln(\frac{\mu}{M_{u}})$, the four running gauge parameters $g_{a}(t), a=1,\dots,4$ all unify to a value $g(0)$ at a scale $M_{u}$. Precision measurements set \cite{Zb,Zd,Ze} 
\begin{equation}
g(0) \simeq .726, \qquad M_{u} \simeq 3 \times 10^{16} GeV  \ .
\label{16}
\end{equation}
Note, $g(0)$ is simply obtained from \cite{Zb} and the equations there in.  
For specificity, we will use these values in our analysis, ignoring as sub-dominant the defocussing effects of possible string thresholds and non-universal soft breaking parameters. The RGEs are given by
\begin{equation}
\frac{dg_{a}}{dt}=\frac{1}{16 {\pi}^{2}} \beta_{a}, \quad a=1,\dots ,4 
\label{17}
\end{equation}
where
\begin{equation}
\beta_{a}=b_{a} g_{a}^{3}, \qquad \vec{b}=(\frac{33}{5}, 1, -3, 12) \ .
\label{18}
\end{equation}
These can be integrated directly to yield
\begin{equation}
g_{a}(t)^{2}= \frac{g(0)^{2}}{1-\frac{g(0)^{2} b_{a} t}{8 {\pi}^{2}}}, \qquad a=1,\dots,4 \ .
\label{19}
\end{equation}

In this section, we are interested in scaling all parameters from the unification mass $M_{u} \simeq 3 \times 10^{16} GeV$ to the $B$-$L$ breaking scale $M_{B-L} \simeq 10^{4} GeV$; that is, 
\begin{equation}
t_{B-L} \simeq -28.7 \leq t \leq 0=t_{u}.
\label{20}
\end{equation}
In this range, we find from (\ref{18}), (\ref{19}) 
that the infrared free parameters $g_{1}, g_{2}$ and $g_{4}$ decrease for small energy-momentum as
\begin{equation}
.441 \leq \frac{g_{1}(t)^{2}}{g(0)^{2}} \leq 1, \ .839 \leq \frac{g_{2}(t)^{2}}{g(0)^{2}} \leq 1, \
.303 \leq \frac{g_{4}(t)^{2}}{g(0)^{2}} \leq 1 
\label{21}
\end{equation}
whereas the asymptotically free coupling $g_{3}$ grows as
\begin{equation}
2.36 \leq \frac{g_{3}(t)^{2}}{g(0)^{2}} \leq 1\ .
\label{22}
\end{equation}

\subsection*{Gaugino Masses:}

Denoting $M_Y=M_{1}$ and $M_{B-L}=M_{4}$, the RGEs for the gaugino masses in (\ref{14}) are \cite{Yb}
\begin{equation}
\frac{dM_{a}}{dt}=\frac{1}{8 {\pi}^{2}} b_{a}g_{a}^{2} M_{a}, \quad a=1,\dots,4 
\label{23}
\end{equation}
where the $b_{a}$ coefficients are given in (\ref{18}). This is immediately solved to give 
\begin{equation}
M_{a}(t)=\frac{M_{a}(0)}{1-\frac{g(0)^{2} b_{a} t}{8 {\pi}^{2}}} \ , \qquad a=1,\dots,4 \ .
\label{24}
\end{equation}
A priori, there is no constraint on the initial values $M_{a}(0)$. In the scaling range (\ref{20}), it follows from (\ref{18}), (\ref{24}) that
\begin{equation}
.441 \leq \frac{M_{1}(t)}{M_{1}(0)} \leq 1, \ .839 \leq \frac{M_{2}(t)}{M_{2}(0)} \leq 1, \
.303 \leq \frac{M_{4}(t)}{M_{4}(0)} \leq 1 
\label{25}
\end{equation}
and
\begin{equation}
2.36 \leq \frac{M_{3}(t)}{M_{3}(0)} \leq 1\ .
\label{26}
\end{equation}

As will be seen shortly, the quantity we will be most interested in is $g_{a}(t)^{2}|M_{a}(t)|^{2}$. Using (\ref{19}) and (\ref{24}), this combination runs under the RG as
\begin{equation}
g_{a}(t)^{2}|M_{a}(t)|^{2}=\frac{g(0)^{2}|M_{a}(0)|^{2}}{(1-\frac{g(0)^{2} b_{a} t}{8 {\pi}^{2}})^{3}}.
\label{g2M2equation}
\end{equation}
Note that even if one assumes that the gaugino masses are ``unified'' at $t=0$, making any ratio $\frac{g_{a}(0)^{2}|M_{a}(0)|^{2}}{g_{b}(0)^{2}|M_{b}(0)|^{2}}$ unity, it is clear that the gluino mass contributions will quickly grow to dominate. For example, at the electroweak scale the ratio of the gluino to the $SU(2)_{L}$ gaugino terms is $25.6$. In this paper, so as to simplify the calculation and allow for a quasi-analytic solution, we will {\it not} assume unified gaugino masses, instead taking 
$|M_{1}(0)|^{2}, |M_{2}(0)|^{2}, |M_{4}(0)|^{2} \ll |M_{3}(0)|^{2}$. It then follows from (\ref{g2M2equation}) that 
\begin{equation}
g_{1}^{2}|M_{1}|^{2}, g_{2}^{2}|M_{2}|^{2}, g_{4}^{2}|M_{4}|^{2} \ll g_{3}^{2}|M_{3}|^{2}
\label{last1}
\end{equation}
over the entire scaling regime.
Recall that ``non-unified'' gaugino masses easily occur in string vacua, while unification requires additional ``minimal'' criteria \cite{Yg, Zc}. These are not generically satisfied in our MSSM theory.

\subsection*{Slepton Masses:}

The RGEs for the slepton  mass parameters $m_{L_{i}}$, $m_{e_{i}}$ and $m_{\nu_{i}}$ are given by \cite{Yb}
\begin{eqnarray}
16{\pi}^{2}\frac{dm_{L_{i}}^{2}}{dt} & = & 2(m_{L_{i}}^{2}+m_{H}^{2}+m_{\nu_{i}}^{2})|\lambda_{\nu_{i}}|^{2}+2(m_{L_{i}}^{2}+m_{\bar{H}}^{2}+m_{e_{i}}^{2})|\lambda_{e_{i}}|^{2} \nonumber \\
                                                              &    & +2|A_{\nu_{i}}|^{2}+2|A_{e_{i}}|^{2}-2g_{1}^{2}|M_{1}|^{2}-6g_{2}^{2}|M_{2}|^{2}-\frac{3}{2}g_{4}^{2}|M_{4}|^{2}  \nonumber \\
                                                              &    & -\frac{3}{5}g_{1}^{2}{\cal{S}}-\frac{3}{4}g_{4}^{2}{\cal{S}'}  \ ,\label{27} \\
16{\pi}^{2}\frac{dm_{e_{i}}^{2}}{dt} & = & 4(m_{L_{i}}^{2}+m_{\bar{H}}^{2}+m_{e_{i}}^{2})|\lambda_{e_{i}}|^{2} +4|A_{e_{i}}|^{2}  \nonumber \\
                                                              &    & -\frac{24}{5}g_{1}^{2}|M_{1}|^{2}-\frac{3}{2}g_{4}^{2}|M_{4}|^{2}  +\frac{6}{5}g_{1}^{2}{\cal{S}}+\frac{3}{4}g_{4}^{2}{\cal{S}'}  \ , \label{28}  \\
16{\pi}^{2}\frac{dm_{\nu_{i}}^{2}}{dt} & = & 4(m_{L_{i}}^{2}+m_{H}^{2}+m_{\nu_{i}}^{2})|\lambda_{\nu_{i}}|^{2} +4|A_{\nu_{i}}|^{2} \nonumber  \\
                                                                 &    & -\frac{3}{2}g_{4}^{2}|M_{4}|^{2}+\frac{3}{4}g_{4}^{2}{\cal{S}'} \label{29} 
\end{eqnarray} 
where
\begin{eqnarray}
{\cal{S}} & =&  m_{H}^{2}-m_{\bar{H}}^{2}+\sum_{i=1}^{3}(m_{Q_{i}}^{2}-2m_{u_{i}}^{2}+m_{d_{i}}^{2}-m_{L_{i}}^{2}+m_{e_{i}}^{2})\nonumber \\ 
  &=&Tr({\frac{\bf Y}{2} {\bf m}^{2}})  ,  \label{30} \\
{\cal{S}'} & =&  \sum_{i=1}^{3}(2m_{Q_{i}}^{2}-m_{u_{i}}^{2}-m_{d_{i}}^{2}-2m_{L_{i}}^{2}+m_{e_{i}}^{2}
+m_{\nu_{i}}^{2}) \nonumber \\ 
  &=&Tr( { {\bf Y}_{B-L} {\bf m}^{2}})  .  \label{31} 
\end{eqnarray}

A full numerical solution of these equations will be presented elsewhere. Here, we give an approximate solution based on the following observations. First, note that the initial conditions for the 
$A$-coefficients in equation (\ref{13}) are, ignoring phenomenological constraints for the time-being, completely arbitrary. However, 
it is conventional \cite{Zc} to let
\begin{equation}
A_{u_{i}}=\lambda_{u_{i}} {\tilde{A}}_{u_{i}}, \ A_{d_{i}}=\lambda_{d_{i}} {\tilde{A}}_{d_{i}},  \ 
A_{\nu_{i}}=\lambda_{\nu_{i}} {\tilde{A}}_{\nu_{i}}, \  A_{e_{i}}=\lambda_{e_{i}} {\tilde{A}}_{e_{i}}
\label{32}
\end{equation}
where the dimensionful $\tilde{A}$-parameters satisfy
\begin{equation}
{\tilde{A}}_{u_{i}} \sim {\cal{O}}(m_{u_{i}}), \ {\tilde{A}}_{d_{i}} \sim {\cal{O}}(m_{d_{i}}), \ {\tilde{A}}_{\nu_{i}} \sim {\cal{O}}(m_{\nu_{i}}), \ {\tilde{A}}_{e_{i}} \sim {\cal{O}}(m_{e_{i}}) \ .
\label{33}
\end{equation}
This is {\it not} a requirement in the ``non-minimal'' string vacua that we are discussing. Be that as it may,
for simplicity of presentation we will assume (\ref{32}) and (\ref{33}) for the remainder of this paper.
Having done this, it follows that every term on the right hand side of equations (\ref{27}), (\ref{28}) and 
(\ref{29}), with the exception of the terms involving the gaugino masses, has the form 
of either $|\lambda |^{2} m^{2}$  or $g^{2} m^{2}$. Our second observation is that the Yukawa couplings
appearing in (\ref{27}), (\ref{28}) and (\ref{29}) satisfy 
\begin{equation}
|\lambda_{\nu_{1}}|  < |\lambda_{\nu_{2}}| < |\lambda_{\nu_{3}}| \simeq 10^{-9} \ll g_{a}, \quad
|\lambda_{e_{1}}|  < |\lambda_{e_{2}}| < |\lambda_{e_{3}}| \simeq 10^{-2} \ll g_{a} 
\label{34}
\end{equation}
throughout  the scaling range (\ref{20}) for $a=1,\dots,4$. Using (\ref{32}), (\ref{33}) and (\ref{34}), it follows that one can approximate the slepton mass RGEs as
\begin{eqnarray}
16{\pi}^{2}\frac{dm_{L_{i}}^{2}}{dt} & \simeq & -2g_{1}^{2}|M_{1}|^{2}-6g_{2}^{2}|M_{2}|^{2}-\frac{3}{2}g_{4}^{2}|M_{4}|^{2}  \nonumber \\
                                                              &    & -\frac{3}{5}g_{1}^{2}{\cal{S}}-\frac{3}{4}g_{4}^{2}{\cal{S}'}  \ ,\label{35} \\
16{\pi}^{2}\frac{dm_{e_{i}}^{2}}{dt} & \simeq & -\frac{24}{5}g_{1}^{2}|M_{1}|^{2}-\frac{3}{2}g_{4}^{2}|M_{4}|^{2}  +\frac{6}{5}g_{1}^{2}{\cal{S}}+\frac{3}{4}g_{4}^{2}{\cal{S}'}  \ , \label{36}  \\
16{\pi}^{2}\frac{dm_{\nu_{i}}^{2}}{dt} & \simeq & -\frac{3}{2}g_{4}^{2}|M_{4}|^{2}+\frac{3}{4}g_{4}^{2}{\cal{S}'} \ . \label{37} 
\end{eqnarray} 
Third, recall that the initial gaugino masses $M_{a}(0), a=1,\dots,4$ are chosen so that (\ref{last1}) is satisfied, but are otherwise arbitrary. Henceforth, we further restrict them so that
\begin{equation}
g_{1}^{2}|M_{1}|^{2}, \ g_{2}^{2}|M_{2}|^{2}, \ g_{4}^{2}|M_{4}|^{2} \ll g_{4}^{2}{\cal{S}'} 
\label{38}
\end{equation}
over the entire scaling range (\ref{20}).  Fourth, we make a specific choice for the scalar masses at the unification scale $M_{u}$. These are taken to be
 \begin{equation}
 m_{H}(0)^{2}=m_{\bar{H}}(0)^{2}, \quad m_{Q_{i}}(0)^{2}=m_{u_{j}}(0)^{2}=m_{d_{k}}(0)^{2}
 \label{39}
 \end{equation}
 and
 \begin{equation}
 m_{L_{i}}(0)^{2}=m_{e_{j}}(0)^{2} \neq m_{\nu_{k}}(0)^{2} \ 
 \label{40}
 \end{equation}
for all $i,j,k=1,2,3$. Note that the sneutrino masses are different than those of the remaining sleptons. {\it This asymmetry is one ingredient in breaking $U(1)_{B-L}$ at an appropriate scale}. Other than that,
 this choice is taken so as to simplify the RGEs as much as possible and to allow a quasi-analytic solution. We point out that soft scalar masses need not be ``universal'' in string theories, since they are not generically ``minimal''. We emphasize that a $B$-$L$/electroweak hierarchy is possible for a much wider range of initial parameters. 
 
 Finally, let us consider the $g_{1}^{2}{\cal{S}}$ and 
 $g_{4}^{2}{\cal{S}'}$ terms. Note from (\ref{30}) and (\ref{31}) that if one chooses the initial scalar masses to satisfy (\ref{39}) and (\ref{40}) then
 \begin{equation}
 {\cal{S}}(0)=0, \quad {\cal{S}'}(0)=\sum_{i=1}^{3}(-m_{L_{i}}(0)^{2}+m_{\nu_{i}}(0)^{2}) \neq 0 \ .
 \label{41}
 \end{equation}
Additionally, we will take ${\cal{S}'} > 0$ with the scale set by the initial sneutrino masses. It then follows from (\ref{41}) that over the scaling range (\ref{20})
\begin{equation}
g_{1}^{2}{\cal{S}} \ll g_{4}^{2}{\cal{S}'} \ .
\label{42}
\end{equation}
In fact, we can go one step further.  Decompose ${\cal{S}'}$ in (\ref{31}) as
\begin{equation}
{\cal{S}'}={\cal{S}'}_{0}+{\cal{S}'}_{1},
\label{43}
\end{equation}
where
\begin{equation}
{\cal{S}'}_{0} =  \sum_{i=1}^{3}(2m_{Q_{i}}^{2}-m_{u_{i}}^{2}-m_{d_{i}}^{2}-m_{L_{i}}^{2}+m_{e_{i}}^{2}), \quad {\cal{S}'}_{1} =  \sum_{i=1}^{3}(-m_{L_{i}}^{2}+m_{\nu_{i}}^{2})  \ .
\label{44}
\end{equation}
Then we see from (\ref{39}) and (\ref{40}) that
\begin{equation}
{\cal{S}'}_{0}(0) = 0, \qquad {\cal{S}'}_{1}(0) =  \sum_{i=1}^{3}(-m_{L_{i}}(0)^{2}+m_{\nu_{i}}(0)^{2}) \neq 0 
\label{45}
\end{equation}
and, hence, over the scaling range (\ref{20})
\begin{equation}
g_{4}^{2}{\cal{S}'}_{0} \ll g_{4}^{2}{\cal{S}'}_{1} \ .
\label{46}
\end{equation}
We conclude from (\ref{38}), (\ref{42}) and (\ref{46}) that a further approximation to the slepton mass RGEs is given by
\begin{eqnarray}
16{\pi}^{2}\frac{dm_{L_{i}}^{2}}{dt} & \simeq & -\frac{3}{4}g_{4}^{2}{\cal{S}'}_{1}  \ ,\label{47} \\
16{\pi}^{2}\frac{dm_{e_{i}}^{2}}{dt} & \simeq & \frac{3}{4}g_{4}^{2}{\cal{S}'}_{1}  \ , \label{48}  \\
16{\pi}^{2}\frac{dm_{\nu_{i}}^{2}}{dt} & \simeq & \frac{3}{4}g_{4}^{2}{\cal{S}'}_{1} \ . \label{49} 
\end{eqnarray} 

From (\ref{44}), (\ref{47}) and (\ref{49}) one finds a RGE for ${\cal{S}'}_{1}$ given by
\begin{equation}
\frac{d{\cal{S}'}_{1}}{dt}= \frac{9}{32 \pi^{2}} g_{4}^{2}{\cal{S}'}_{1} \ .
\label{50}
\end{equation}
Using (\ref{19}), this is easily solved to give
\begin{equation}
{\cal{S}'}_{1}(t)= \frac{{\cal{S}'}_{1}(0)}
{(1-\frac{g(0)^{2}b_{4}t}{8 {\pi^{2}}})^{\frac{9}{4b_{4}}} } \ .
\label{51}
\end{equation}
It follows from (\ref{16}) and (\ref{18}) that in the scaling range (\ref{20})
\begin{equation}
.800 \leq \frac{{\cal{S}'}_{1}(t)}{{\cal{S}'}_{1}(0)} \leq 1 \ .
\label{52}
\end{equation}
It is now straightforward to solve (\ref{47}), (\ref{48}) and (\ref{49}) for $m_{L_{i}}^{2}$, $m_{e_{i}}^{2}$ and $m_{\nu_{i}}^{2}$ respectively.  From (\ref{19}), (\ref{48}) and (\ref{51}) one finds
\begin{equation}
m_{L_{i}}(t)^{2}=m_{L_{i}}(0)^{2}+\frac{1}{6}(1-(1-\frac{g(0)^{2}b_{4}t}{8 {\pi}^{2}})^{-9/4b_{4}})
{\cal{S}'}_{1}(0)
\label{53}
\end{equation}
and
\begin{eqnarray}
m_{e_{i}}(t)^{2} & = & m_{e_{i}}(0)^{2}-\frac{1}{6}(1-(1-\frac{g(0)^{2}b_{4}t}{8 {\pi}^{2}})^{-9/4b_{4}})
{\cal{S}'}_{1}(0) \label{54} \\
m_{\nu_{i}}(t)^{2} & = & m_{\nu_{i}}(0)^{2}-\frac{1}{6}(1-(1-\frac{g(0)^{2}b_{4}t}{8 {\pi}^{2}})^{-9/4b_{4}})
{\cal{S}'}_{1}(0)  \label{55} \ .
\end{eqnarray}
These equations are almost identical with the exception of the sign of the second term. This is positive for $m_{L_{i}}^{2}$, whereas it is negative in the expressions for $m_{e_{i}}^{2}$ and $m_{\nu_{i}}^{2}$.
This sign has important physical consequences. Using (\ref{16}) and (\ref{18}), the mass parameter parameter for $m_{L_{i}}^{2}$ increases from its initial value at $t=0$ to
\begin{equation}
m_{L_{i}}(t_{B-L})^{2}=m_{L_{i}}(0)^{2}+(3.35 \times 10^{-2}){\cal{S}'}_{1}(0) 
\label{56}
\end{equation}
at $t_{B-L}$. Thus it always remains positive. On the other hand, the mass parameters for $m_{e_{i}}^{2}$ and $m_{\nu_{i}}^{2}$ start at their initial values at $t=0$ but decrease to
\begin{eqnarray}
m_{e_{i}}(t_{B-L})^{2} & = & m_{e_{i}}(0)^{2}-(3.35 \times 10^{-2}){\cal{S}'}_{1}(0) \ , \label{57} \\
m_{\nu_{i}}(t_{B-L})^{2} & = & m_{\nu_{i}}(0)^{2}-(3.35 \times 10^{-2}){\cal{S}'}_{1}(0) \label{58}
\end{eqnarray}
at $t_{B-L}$. If the initial masses are sufficiently small and ${\cal{S}'}_{1}(0)$ is sufficiently large, then one or more of these squared masses can become negative signaling possible symmetry breaking.

\subsection*{ Spontaneous $B$-$L$ Breaking:}

Given the RGE solutions in the previous subsections, one can now study the scalar field potential at any scale in the range (\ref{20}) and, in particular, search for local minima. Here, we limit the discussion to the slepton fields. Higgs fields and squarks will be discussed in the next subsection, where the consistency of this two-step procedure will be demonstrated. 

Let us begin by considering the quadratic mass terms near the origin of field space. The relevant  part of the scalar potential is
\begin{equation}
V=V_{2s} + \frac{1}{2} D_{B-L}^{2}
\label{73}
\end{equation}
where $V_{2s}$ and $D_{B-L}$ are given in (\ref{12}) and (\ref{10}) respectively. Expanding this using the $B$-$L$ quantum numbers listed in (\ref{4}) , the quadratic terms are given at any scale $t$ by
\begin{equation}
V=\dots+\sum_{i=1}^{3} (m_{L_{i}\rm }^{2}|L_{i}|^{2}+m_{e_{i}\rm  }^{2}|e_{i}|^{2}+m_{\nu_{i}\rm  }^{2}|\nu_{i}|^{2})+\dots 
\label{74}
\end{equation}
The reader should recall that the FI terms have been absorbed into the soft mass parameters.  All $\beta$ functions in the D-eliminated formalism are written in terms of these redefined masses.  As the theory is scaled from $t=0$ toward $t_{B-L}$, the $m_{L_{i}}^{2}$, $m_{e_{i}}^{2}$ and $m_{\nu_{i}}^{2}$ parameters scale as in (\ref{56}),(\ref{57}) and (\ref{58}) respectively.

The first requirement for spontaneous $B$-$L$ breaking is that at least one of the slepton squared masses becomes negative at $t_{B-L}$. Clearly, this cannot happen for $m_{L_{i}\rm }(t_{B-L})^{2}$, which is always positive. However, if the initial squared masses are sufficiently small and ${\cal{S}'}_{1}(0)$ sufficiently large, both $m_{e_{i}\rm}(t_{B-L})^{2}$ and $m_{\nu_{i}\rm }(t_{B-L})^{2}$ can become negative. Since the $e_{i}$ fields are electrically charged, we do not want them to get a vacuum expectation value and, hence, we want  $m_{e_{i}\rm}(t_{B-L})^{2}$ to be positive. On the other hand, the $\nu_{i}$ fields are neutral in all quantum numbers except $B$-$L$. Hence, if they get a nonzero VEV this will spontaneously break $B$-$L$ at $t_{B-L}$, but leave the $SU(3)_{C} \times SU(2)_{L} \times U(1)_{Y}$ gauge symmetry unbroken. We now show that for a wide range of initial parameters this is indeed possible. For simplicity, let us choose the initial right-handed slepton masses to be
\begin{equation}
m_{\nu_{1}}(0)=m_{\nu_{2}}(0)=Cm_{\nu}(0), \quad m_{\nu_{3}}(0)=m_{\nu}(0)
\label{80}
\end{equation}
 and
\begin{equation}
m_{e_{1}}(0)=m_{e_{2}}(0)=m_{e_{3}}(0)=Am_{\nu}(0)
\label{81}
\end{equation}
for some dimensionless constants $C$ and $A$ to be determined. Using (\ref{40}), (\ref{41}) 
and (\ref{80}), (\ref{81}) we see that ${\cal{S}'}_{1}(0)$ is parameterized  by 
\begin{equation}
{\cal{S}'}_{1}(0)=(1+2C^{2}-3A^{2})m_{\nu}(0)^{2} \ .
\label{82}
\end{equation}
Let us first consider $\nu_{3}$. It follows from (\ref{58}), (\ref{80}) and (\ref{82}) that
\begin{equation}
m_{\nu_{3}\rm }(t_{B-L})^{2}=(1-(3.35 \times 10^{-2})(1+2C^{2}-3A^{2}))m_{\nu}(0)^{2} \ .
\label{83}
\end{equation}
For specificity, we will henceforth take
\begin{equation}
(3.35 \times 10^{-2})(1+2C^{2}-3A^{2})=5 \ .
\label{84}
\end{equation}
This choice yields the simple result that 
\begin{equation}
m_{\nu_{3}\rm }(t_{B-L})^{2}=-4 m_{\nu}(0)^{2} \ ,
\label{85}
\end{equation}
suggesting a non-zero VEV in the $\nu_{3}$ direction. Now consider $\nu_{1}$, $\nu_{2}$. In this case, we see from (\ref{58}), (\ref{80}), (\ref{82}) and (\ref{84}) that
\begin{equation}
m_{\nu_{1,2}\rm }(t_{B-L})^{2}=(C^{2}-5)m_{\nu}(0)^{2} \ .
\label{86}
\end{equation}
Now let us examine the $e_{1,2,3}$ fields. Using (\ref{57}), (\ref{81}), (\ref{82}) and (\ref{84}) one finds
\begin{equation}
m_{e_{i}\rm }(t_{B-L})^{2}=(A^{2}-5)m_{\nu}(0)^{2} 
\label{87}
\end{equation}
for $i=1,2,3$. Since all $m_{e_{i}\rm}(t_{B-L})^{2}$ must be positive, the coefficient $A$ must satisfy $A^{2}-5 > 0$. Again, for specificity we will choose
\begin{equation}
A=\sqrt{6} \ ,
\label{88}
\end{equation}
which yields the simple result that
\begin{equation}
m_{e_{i}\rm }(t_{B-L})^{2}=m_{\nu}(0)^{2} 
\label{89}
\end{equation}
for $i=1,2,3$. Putting $A=\sqrt{6}$ into expression (\ref{84}) gives 
\begin{equation}
C=9.12 \ .
\label{90}
\end{equation}
Hence, we see from (\ref{86}) that both $m_{\nu_{1,2}\rm }(t_{B-L})^{2}$ are positive and given by
\begin{equation}
m_{\nu_{1,2}\rm }(t_{B-L})^{2}=78.2\ m_{\nu}(0)^{2} \ .
\label{91}
\end{equation}
We conclude from (\ref{89}) and (\ref{91}) that, near the origin of field space, there are positive quadratic mass terms in the  $e_{1,2,3}$ and $\nu_{1,2}$ field directions. Finally, let us consider the 
$L_{i}, i=1,2,3$ masses. Using (\ref{40}), (\ref{56}), (\ref{81}), (\ref{82}), (\ref{84}) and (\ref{88}), we find that 
\begin{equation}
m_{L_{1,2,3}\rm }(t_{B-L})^{2}=11 \ m_{\nu}(0)^{2} \ ,
\label{92}
\end{equation}
that is, a positive quadratic mass term in each if the $L_{i}$ field directions at the origin. We note for future reference that using (\ref{88}) and (\ref{90}),  equation (\ref{82}) becomes
\begin{equation}
{\cal{S}'}_{1}(0)=149 \ m_{\nu}(0)^{2} \ . 
\label{93}
\end{equation}

Given these mass terms, as well as $g_{4}$ in (\ref{21}), one can minimize the complete potential to determine the vacuum at $t_{B-L}$. The part of this potential relevant to finding a local minimum in the slepton fields is $V$ in (\ref{73}). We note that, in principle, the $|\lambda_{\nu_{3}}|^{2}|L_{3}|^{2}|\nu_{3}|^{2}$ term in the $|\partial_{H}W|^{2}$ contribution to the potential, where $W$ is given in (\ref{6}), could play a role in selecting the vacuum. However, since $\lambda_{\nu_{3}} \ll g_{4}$, this term can be safely ignored.
Expanding out the contributions to $V$ and collecting terms, the important interactions  are
\begin{eqnarray}
V & = & m_{\nu_{3}\rm}^{2}|\nu_{3}|^{2}+\frac{\frac{3}{4}g_{4}^{2}}{2}|\nu_{3}|^{4} +(m_{\nu_{1,2}\rm}^{2}
+\frac{3}{4}g_{4}^{2}|\nu_{3}|^{2})|\nu_{1,2}|^{2} \nonumber \\
&  &+\sum_{i=1}^{3} ((m_{e_{i}\rm}^{2}+\frac{3}{4}g_{4}^{2}|\nu_{3}|^{2})|e_{i}|^{2}+(m_{L_{i}\rm}^{2}-\frac{3}{4}g_{4}^{2}|\nu_{3}|^{2})|L_{i}|^{2}) +\dots  \label{94}
\end{eqnarray}
Recalling from (\ref{85}) that $m_{\nu_{3}\rm}(t_{B-L})^{2}=-4m_{\nu}(0)^{2}$, the first two terms in (\ref{94}) can be minimized by the non-zero VEV
\begin{equation}
\langle \nu_{3} \rangle= \frac{2 \ m_{\nu}(0)}{\sqrt{\frac{3}{4}}g_{4}(t_{B-L})} \ .
\label{95}
\end{equation}
Is this point a local minimum of the complete slepton potential? To determine this, consider potential 
(\ref{94}) near this VEV. Clearly, the first derivatives of all fields vanish at this point. Now determine the masses in all slepton field directions evaluated at (\ref{95}). It follows from (\ref{94}) that
\begin{equation}
V= \langle m_{\nu_{3}}^{2} \rangle |\delta \nu_{3}|^{2} +\langle m_{\nu_{1,2}}^{2} \rangle |\nu_{1,2}|^{2}
+\sum_{i=1}^{3} (\langle m_{e_{i}}^{2} \rangle|e_{i}|^{2}+\langle m_{L_{i}}^{2} \rangle |L_{i}|^{2}) + \dots \ ,
\label{96}
\end{equation}
where
\begin{eqnarray}
\langle m_{\nu_{3}}^{2} \rangle & = & -2 \ m_{\nu_{3}\rm}^{2}, \qquad \  \ \
\quad \langle m_{\nu_{1,2}}^{2} \rangle = m_{\nu_{1,2}\rm}^{2}+\frac{3}{4}g_{4}^{2}\langle \nu_{3} \rangle^{2},
\nonumber \\
\langle m_{e_{i}}^{2} \rangle & = & m_{e_{i}\rm}^{2}+\frac{3}{4}g_{4}^{2} \langle \nu_{3} \rangle^{2},  
\quad \langle m_{L_{i}}^{2} \rangle  =  m_{L_{i}\rm}^{2}-\frac{3}{4}g_{4}^{2}\langle \nu_{3} \rangle^{2}
\label{97} \ .
\end{eqnarray}
Using (\ref{85}), (\ref{89}), (\ref{91}), (\ref{92}) and (\ref{95}) we find that at $t_{B-L}$
\begin{eqnarray}
\langle m_{\nu_{3}}^{2} \rangle & = & 8 \ m_{\nu}(0)^{2}, \quad \langle m_{\nu_{1,2}}^{2} \rangle = 82.2 \ m_{\nu}(0)^{2},  \nonumber \\
\langle m_{e_{i}}^{2} \rangle & = & 5 \ m_{\nu}(0)^{2}, \quad  \langle m_{L_{i}}^{2} \rangle  =  7 \ m_{\nu}(0)^{2} \ .
\label{98}
\end{eqnarray}
Since all of these masses are positive, we conclude that the vacuum specified by
\begin{equation}
\langle \nu_{1,2} \rangle= 0, \quad \langle \nu_{3} \rangle= \frac{2 \ m_{\nu}(0)}{\sqrt{\frac{3}{4}}g_{4}(t_{B-L})}
\label{99}
\end{equation}
and
\begin{equation}
\langle e_{i} \rangle= \langle L_{i} \rangle= 0 \quad i=1,2,3
\label{100}
\end{equation}
is indeed a local minimum of the slepton potential $V$. This vacuum spontaneously breaks the gauged $B-L$ symmetry while preserving the remaining $SU(3)_{C} \times SU(2)_{L} \times U(1)_{Y}$ gauge group. The massless Goldstone boson is ``eaten'' by the  $B$-$L$ vector boson giving it a mass 
\begin{equation}
M_{A_{B-L}}=\sqrt{2} \ g_{B-L}(t_{B-L}) \ \langle \nu_{3} \rangle \ .
\label{yyy1}
\end{equation}
Using (\ref{15}) and (\ref{99}), this becomes
\begin{equation}
M_{A_{B-L}}= 2\sqrt{2} \ m_{\nu}(0) \ .
\label{101}
\end{equation}

We now have to  include the Higgs fields and squarks, analyze their masses at $t_{B-L}$ around vacuum (\ref{99}), (\ref{100}) and show that this is a local minimum in the complete scalar field space. To do this, one must discuss the Higgs and squark RGEs, to which we now turn.

\section{The RGEs for Higgs Fields and Squarks}

In this section, we present and analyze the RGEs for the up and down Higgs fields, the left and right squarks, as well as the associated $\mu$ and $B$ parameters. Using these results, we compute  the Higgs and squark masses at $t_{B-L}$ and show that they are all positive at vacuum (\ref{99}), (\ref{100}), as desired.

\subsection*{``Up'' Higgs and Squark Masses:}

The RGEs for the ``up'' Higgs and squark mass parameters $m_{H}$, $m_{Q_{i}}$ and $m_{u_{i}}$ are given by \cite{Yb}
\begin{eqnarray}
16{\pi}^{2}\frac{dm_{H}^{2}}{dt} & = & {\sum_{i=1}^{3}}(6(m_{Q_{i}}^{2}+m_{H}^{2}+m_{u_{i}}^{2})|\lambda_{u_{i}}|^{2}+2(m_{L_{i}}^{2}+m_{H}^{2}+m_{\nu_{i}}^{2})|\lambda_{\nu_{i}}|^{2} \nonumber \\
                                                              &    & +6|A_{u_{i}}|^{2}+2|A_{\nu_{i}}|^{2})-\frac{6}{5}g_{1}^{2}|M_{1}|^{2}-6g_{2}^{2}|M_{2}|^{2} \label{106}  \\
                                                              &    & +\frac{3}{5}g_{1}^{2}{\cal{S}}, \nonumber \\
16{\pi}^{2}\frac{dm_{Q_{i}}^{2}}{dt} & = & 2(m_{Q_{i}}^{2}+m_{H}^{2}+m_{u_{i}}^{2})|\lambda_{u_{i}}|^{2} +2(m_{Q_{i}}^{2}+m_{\bar{H}}^{2}+m_{d_{i}}^{2})|\lambda_{d_{i}}|^{2} \nonumber \\
                                                              &    &  +2|A_{u_{i}}|^{2} +2|A_{d_{i}}|^{2} -\frac{12}{15}g_{1}^{2}|M_{1}|^{2} -6g_{2}^{2}|M_{2}|^{2} -\frac{32}{3}g_{3}^{2}|M_{3}|^{2}  \nonumber \\
                                                              &    & -\frac{1}{6}g_{4}^{2}|M_{4}|^{2}+\frac{1}{5}g_{1}^{2}{\cal{S}}+\frac{1}{4}g_{4}^{2}{\cal{S}'}  \ , \label{107}  \\
16{\pi}^{2}\frac{dm_{u_{i}}^{2}}{dt} & = & 4(m_{Q_{i}}^{2}+m_{H}^{2}+m_{u_{i}}^{2})|\lambda_{u_{i}}|^{2} +4|A_{u_{i}}|^{2} \nonumber  \\
                                                                 &    &  -\frac{32}{15}g_{1}^{2}|M_{1}|^{2}  -\frac{32}{3}g_{3}^{2}|M_{3}|^{2} -\frac{1}{6}g_{4}^{2}|M_{4}|^{2}  \label{108}  \\
                                                                 &    & -\frac{4}{5}g_{1}^{2}{\cal{S}} -\frac{1}{4}g_{4}^{2}{\cal{S}'}  \nonumber                                                       
\end{eqnarray}
where ${\cal{S}}$ and ${\cal{S}'}$ are given in (\ref{30}) and (\ref{31}) respectively.

A full numerical solution of these equations will be presented elsewhere. Here, we give an approximate solution based on the following observations. First, as discussed in the previous section we will assume that the $A$-coefficients satisfy (\ref{32}) and (\ref{33}). Having done this, it follows that every term on the right hand side of equations (\ref{106}), (\ref{107}) and 
(\ref{108}), with the exception of the terms involving the gaugino masses, has the form 
of either $|\lambda |^{2} m^{2}$  or $g^{2} m^{2}$. Our second observation is that the Yukawa couplings
appearing in (\ref{106}), (\ref{107}) and (\ref{108}) satisfy 
\begin{eqnarray}
|\lambda_{\nu_{1}}|  < |\lambda_{\nu_{2}}| < |\lambda_{\nu_{3}}|& \simeq& 10^{-9} \ll g_{a}, \nonumber \\
|\lambda_{d_{1}}|  < |\lambda_{d_{2}}| < |\lambda_{d_{3}}| &\simeq& 5 \times 10^{-2} \ll g_{a} , 
\label{109} 
\end{eqnarray}
and
\begin{equation}
|\lambda_{u_{1}}|  < |\lambda_{u_{2}}| \simeq 10^{-2} \ll g_{a}, \quad |\lambda_{u_{3}}| \simeq 1
\label{110}
\end{equation}
throughout  the scaling range (\ref{20}) for $a=1,\dots,4$. We see from (\ref{110}) that 
{\it $|\lambda_{u_{3}}|$ is of ${\cal{O}}(1)$ and, hence, terms containing it cannot be dropped from the RGEs}. As in the previous section, we will continue to assume that the initial gaugino masses $M_{a}(0), a=1,2,4$ are chosen sufficiently small that the inequalities in (\ref{38}) remain satisfied. However, {\it we will not make a similar assumption about $M_{3}(0)$, allowing it, for the time being, to be arbitrarily large and not sub-leading to ${\cal{S'}}_{1}$ }. Since the initial scalar masses satisfy (\ref{39}) and (\ref{40}), the inequalities  (\ref{42}) and (\ref{46}) are still satisfied. Finally, we will assume that terms with $g_{1}^{2}{\cal{S}}$ are small compared to any term proportional to $|\lambda_{u_{3}}|^{2}$. Using all these inputs and assumptions, it follows that one can approximate the squark mass RGEs as
\begin{eqnarray}
 16{\pi}^{2}\frac{dm_{Q_{3}}^{2}}{dt} & \simeq & 2(m_{Q_{3}}^{2}+m_{H}^{2}+m_{u_{3}}^{2})|\lambda_{u_{3}}|^{2} +2|\lambda_{u_{3}}|^{2}|{\tilde{A}}_{u_{3}}|^{2}  \nonumber \\
     &    & -\frac{32}{3}g_{3}^{2}|M_{3}|^{2} +\frac{1}{4}g_{4}^{2}{\cal{S}'}_{1}  \ , \label{112}  \\
16{\pi}^{2}\frac{dm_{u_{3}}^{2}}{dt} & \simeq & 4(m_{Q_{3}}^{2}+m_{H}^{2}+m_{u_{3}}^{2})|\lambda_{u_{3}}|^{2} +4|\lambda_{u_{3}}|^{2}|{\tilde{A}}_{u_{3}}|^{2} \nonumber  \\
    &    & -\frac{32}{3}g_{3}^{2}|M_{3}|^{2} -\frac{1}{4}g_{4}^{2}{\cal{S}'}_{1}  \label{113}                                                     
\end{eqnarray}
and 
\begin{eqnarray}
 16{\pi}^{2}\frac{dm_{Q_{1,2}}^{2}}{dt} & \simeq & -\frac{32}{3}g_{3}^{2}|M_{3}|^{2} +\frac{1}{4}g_{4}^{2}{\cal{S}'}_{1}  \ , \label{114}  \\
16{\pi}^{2}\frac{dm_{u_{1,2}}^{2}}{dt} & \simeq & -\frac{32}{3}g_{3}^{2}|M_{3}|^{2} -\frac{1}{4}g_{4}^{2}{\cal{S}'}_{1} \ . \label{115}                                                     
\end{eqnarray}

Let us begin by analyzing equations (\ref{112}) and (\ref{113}). Note that (\ref{112}) and (\ref{113}) can be written as
\begin{eqnarray}
 16{\pi}^{2}\frac{dm_{Q_{3}}^{2}}{dt} & \simeq & 16{\pi}^{2}\frac{d(\frac{1}{3}m_{H}^{2})}{dt}  -\frac{32}{3}g_{3}^{2}|M_{3}|^{2} +\frac{1}{4}g_{4}^{2}{\cal{S}'}_{1}  \ , \label{116}  \\
16{\pi}^{2}\frac{dm_{u_{3}}^{2}}{dt} & \simeq & 16{\pi}^{2}\frac{d(\frac{2}{3}m_{H}^{2})}{dt} -\frac{32}{3}
g_{3}^{2}|M_{3}|^{2} -\frac{1}{4}g_{4}^{2}{\cal{S}'}_{1}  \label{117}                                                     
\end{eqnarray}
respectively. These are easily solved to give
\begin{eqnarray}
m_{Q_{3}}^{2} & \simeq & \frac{1}{3}m_{H}^{2}-\frac{2}{3 {\pi}^{2}} \int_{0}^{t}{g_{3}^{2}|M_{3}|^{2}}+\frac{1}{64 {\pi}^{2}} \int_{0}^{t}g_{4}^{2}{\cal{S}'}_{1}+{\cal{C}}_{Q_{3}} \ , 
  \label{118} \\
m_{u_{3}}^{2} & \simeq & \frac{2}{3}m_{H}^{2}-\frac{2}{3 {\pi}^{2}} \int_{0}^{t}{g_{3}^{2}|M_{3}|^{2}}-\frac{1}{64 {\pi}^{2}} \int_{0}^{t}{g_{4}^{2}{\cal{S}'}_{1}} +{\cal{C}}_{u_{3}}   \label{119} 
\end{eqnarray}
where 
\begin{eqnarray}
-\frac{2}{3 {\pi}^{2}} \int_{0}^{t}{g_{3}^{2}|M_{3}|^{2}} & = & -\frac{8}{3b_{3}}( \frac{1}{(1-\frac{g(0)^{2}b_{3}t}{8 {\pi}^{2}})^{2}}-1)|M_{3}(0)|^{2} \ , \label{120} \\
-\frac{1}{64 {\pi}^{2}} \int_{0}^{t}{g_{4}^{2}{\cal{S}'}_{1}} & = & -\frac{1}{18}( \frac{1}{(1-\frac{g(0)^{2}b_{4}t}{8 {\pi}^{2}})^{\frac{9}{4b_{4}}}}-1){\cal{S}'}_{1}(0)  \label{121}
\end{eqnarray}
are evaluated using (\ref{19}), (\ref{24}) and (\ref{51}). The integration constants are
\begin{equation}
{\cal{C}}_{Q_{3}}=m_{Q_{3}}(0)^{2}-\frac{1}{3}m_{H}(0)^{2}, \quad {\cal{C}}_{u_{3}}=m_{Q_{3}}(0)^{2}-\frac{2}{3}m_{H}(0)^{2} 
\label{122}
\end{equation}
where we have used our assumption (\ref{39}) that the initial squark masses are degenerate. Note that 
since $b_{3}=-3$ and $t \leq 0$, expression (\ref{120}) is always non-negative. 
Similarly, since $b_{4}=12$, it follows that (\ref{121}) is also non-negative. 

Now consider the $m_{H}^{2}$ equation (\ref{106}). If we insert (\ref{118}) and (\ref{119}) into (\ref{106}), we get an expression for $m_{H}^{2}$ without the the squark mass squared terms.  It is important to note that in doing so, we would gain a term in the beta function that depends on the gluino mass squared.  If we then apply (\ref{38}) and the discussion above (\ref{112}) (where we took the gluino mass squared term to be of the same order as $ {\cal{S}'}_{1} $ in our approximation), we get 
\begin{equation}
\frac{dm_{H}^{2}}{dt} \simeq  \frac{3}{8{\pi}^{2}}|\lambda_{u_{3}}|^{2}(2m_{H}^{2}-\frac{4}{3 {\pi}^{2}} 
\int_{0}^{t}{g_{3}^{2}|M_{3}|^{2}} +m_{{\cal{C}}}^{2}+|{\tilde{A}}_{u_{3}}|^{2} )
 \label{123}
 \end{equation}
with $m_{{\cal{C}}}^{2}$ defined by
\begin{equation}
m_{{\cal{C}}}^{2}= {\cal{C}}_{Q_{3}} + {\cal{C}}_{u_{3}}= 2m_{Q_{3}}(0)^{2}-m_{H}(0)^{2} \ .
\label{124}
\end{equation}
Henceforth, to simplify our discussion, we assume 
\begin{equation}
m_{Q_{3}}(0)^{2} = \frac{m_{H}(0)^{2}}{2} \ ,
\label{125}
\end{equation}
which sets
\begin{equation}
m_{{\cal{C}}}^{2}=0 \ .
\label{125A}
\end{equation}
 Furthermore, we will choose the initial value ${\tilde{A}}_{u_{3}}(0)$ so that
 \begin{equation}
|{\tilde{A}}_{u_{3}}|^{2} \ll 2m_{H}^{2}-\frac{4}{3 {\pi}^{2}} 
\int_{0}^{t}{g_{3}^{2}|M_{3}|^{2}}
 \label{128}
 \end{equation}
over the entire scaling range. As a final simplification, we note that $\lambda_{u_{3}}$ scales by only a few percent over the scaling range. Hence, we approximate it as a constant with its phenomenological value of
\begin{equation}
\lambda_{u_{3}} \simeq 1 \ .
\label{buddy1}
\end{equation}
Using (\ref{125A}), (\ref{128})  and (\ref{buddy1}), RGE equation (\ref{123}) simplifies to
\begin{equation}
\frac{dm_{H}^{2}}{dt} \simeq  \frac{3}{8{\pi}^{2}}(2m_{H}^{2}-\frac{4}{3 {\pi}^{2}} 
\int_{0}^{t}{g_{3}^{2}|M_{3}|^{2}}) \ .
 \label{123A}
 \end{equation}
In this paper, we will solve equation (\ref{123A}) subject to the constraint that $m_{H}^{2}$, which is positive at $t=0$, remain positive over the entire scaling range. 
Then, RGE (\ref{123A}) is equivalent to an integral equation for $m_{H}^{2}$ given by 
\begin{equation}
m_{H}^{2} \simeq m_{H}(0)^{2} 
e^{  - \frac{3}{4{\pi}^{2}} \int_{t}^{0}      
(1  +[ \frac{-\frac{2}{3 {\pi}^{2}}   
\int_{0}^{t'}{g_{3}^{2}|M_{3}|^{2}}
}{m_{H}^{2}} ] ) } \ .
\label{129}
\end{equation}

Note that (\ref{125}) implies
\begin{equation}
{\cal{C}}_{Q_{3}}=\frac{1}{6}m_{H}(0)^{2}, \quad {\cal{C}}_{u_{3}}=-\frac{1}{6}m_{H}(0)^{2} \ .
\label{126}
\end{equation}
It follows that the $m_{Q_{3}}^{2}$ and $m_{u_{3}}^{2}$ equations in (\ref{118}) and (\ref{119}) become
\begin{eqnarray}
m_{Q_{3}}^{2} & \simeq & \frac{1}{3}m_{H}^{2}-\frac{2}{3 {\pi}^{2}} \int_{0}^{t}{g_{3}^{2}|M_{3}|^{2}}+\frac{1}{64 {\pi}^{2}} \int_{0}^{t}g_{4}^{2}{\cal{S}'}_{1}+\frac{1}{6}m_{H}(0)^{2} \ , 
  \label{118A} \\
m_{u_{3}}^{2} & \simeq & \frac{2}{3}m_{H}^{2}-\frac{2}{3 {\pi}^{2}} \int_{0}^{t}{g_{3}^{2}|M_{3}|^{2}}-\frac{1}{64 {\pi}^{2}} \int_{0}^{t}{g_{4}^{2}{\cal{S}'}_{1}} -\frac{1}{6}m_{H}(0)^{2}   \label{119A} 
\end{eqnarray}
respectively.
Finally, let us consider RGEs (\ref{114}) and (\ref{115}). These are easily solved to give
\begin{eqnarray}
m_{Q_{1,2}}^{2} & \simeq & -\frac{2}{3 {\pi}^{2}} \int_{0}^{t}{g_{3}^{2}|M_{3}|^{2}}
+\frac{1}{64 {\pi}^{2}} \int_{0}^{t}{g_{4}^{2}{\cal{S}'}_{1}} +\frac{m_{H}(0)^{2}}{2} \ , \label{130} \\
m_{u_{1,2}}^{2} & \simeq & -\frac{2}{3 {\pi}^{2}} \int_{0}^{t}{g_{3}^{2}|M_{3}|^{2}}
-\frac{1}{64 {\pi}^{2}} \int_{0}^{t}{g_{4}^{2}{\cal{S}'}_{1}} +\frac{m_{H}(0)^{2}}{2}  \label{131} 
\end{eqnarray}
where the $g_{3}^{2}|M_{3}|^{2}$ and $g_{4}^{4}{\cal{S}'}_{1}$ integrals are given in (\ref{120}) and (\ref{121}) respectively and we have used assumption (\ref{39}) that the initial squark masses are degenerate.

\subsection*{``Down'' Higgs and Squark Masses:}

The RGEs for the ``down'' Higgs and squark mass parameters $m_{\bar{H}}$ and $m_{d_{i}}$ are given by \cite{Yb}
\begin{eqnarray}
16{\pi}^{2}\frac{dm_{\bar{H}}^{2}}{dt} & = & {\sum_{i=1}^{3}}(6(m_{Q_{i}}^{2}+m_{\bar{H}}^{2}+m_{d_{i}}^{2})|\lambda_{d_{i}}|^{2}+2(m_{L_{i}}^{2}+m_{\bar{H}}^{2}+m_{e_{i}}^{2})|\lambda_{e_{i}}|^{2} \nonumber \\
                                                              &    & +6|A_{d_{i}}|^{2}+2|A_{e_{i}}|^{2})-\frac{6}{5}g_{1}^{2}|M_{1}|^{2}-6g_{2}^{2}|M_{2}|^{2} \label{135}  \\
                                                              &    & -\frac{3}{5}g_{1}^{2}{\cal{S}}, \nonumber \\
16{\pi}^{2}\frac{dm_{d_{i}}^{2}}{dt} & = & 4(m_{Q_{i}}^{2}+m_{\bar{H}}^{2}+m_{d_{i}}^{2})|\lambda_{d_{i}}|^{2} +4|A_{d_{i}}|^{2} \nonumber  \\
                                                                 &    &  -\frac{8}{15}g_{1}^{2}|M_{1}|^{2}  -\frac{32}{3}g_{3}^{2}|M_{3}|^{2} -\frac{1}{6}g_{4}^{2}|M_{4}|^{2}  \label{136}  \\
                                                                 &    & +\frac{2}{5}g_{1}^{2}{\cal{S}} -\frac{1}{4}g_{4}^{2}{\cal{S}'}  \nonumber                                                       
\end{eqnarray}
where ${\cal{S}}$ and ${\cal{S}'}$ are given in (\ref{30}) and (\ref{31}) respectively. First consider the $m_{d_{i}}$ equation. Using the assumptions listed in (\ref{32}), (\ref{33}), (\ref{38}), (\ref{46})
and (\ref{109}), equation (\ref{136}) can be approximated by
\begin{equation}
16{\pi}^{2}\frac{dm_{d_{i}}^{2}}{dt}  \simeq  -\frac{32}{3}g_{3}^{2}|M_{3}|^{2}-\frac{1}{4}g_{4}^{2}{\cal{S}'}_{1} \ . 
\label{137}
\end{equation}
This can be immediately integrated to give
\begin{equation}
m_{d_{i}}^{2} \simeq -\frac{2}{3 {\pi}^{2}} \int_{0}^{t}{g_{3}^{2}|M_{3}|^{2}}
-\frac{1}{64 {\pi}^{2}} \int_{0}^{t}{g_{4}^{2}{\cal{S}'}_{1}}+\frac{m_{H}(0)^{2}}{2} \ ,
\label{138}
\end{equation}
where the first and second terms on the right hand side are evaluated in (\ref{120}) and (\ref{121}) respectively, and we have used assumptions (\ref{39}) and (\ref{125}). Note that these are both non-negative and, hence, $m_{d_{i}}^{2}, i=1,2,3$ are all non-negative throughout the entire scaling range (\ref{20}). Now consider the $m_{\bar{H}}$ equation (\ref{135}). All of the terms on the right hand side of this expression are small compared to $g_{3}^{2}|M_{3}|^{2}$ and $g_{4}^{2}{\cal{S}'}_{1}$.  In addition, these terms are multiplied by $|\lambda_{d_{i}}|^{2}$, which is small.  Hence, to the order we are working
\begin{equation}
\frac{dm_{\bar{H}}^{2}}{dt} \simeq 0
\label{139}
\end{equation}
and
\begin{equation}
m_{\bar{H}}^{2} \simeq m_{H}(0)^{2} \ ,
\label{140}
\end{equation}
where we have used (\ref{39}).

\subsection*{The $\mu$ Parameter:}

The $\mu$ parameter enters the potential for the Higgs supermultiplets $H$, $\bar{H}$
through the superpotential $W$ in (\ref{6}). The RGE is given by \cite{Yb}
\begin{equation}
16 {\pi}^{2} \frac{d\mu}{dt}=\mu \left(\sum_{i=1}^{3} (3|\lambda_{u_{i}}|^{2}+3|\lambda_{d_{i}}|^{2}+|\lambda_{\nu_{i}}|^{2}+|\lambda_{e_{i}}|^{2}) -\frac{3}{5}g_{1}^{2}-3g_{2}^{2}\right) \ .
\label{102}
\end{equation}
This is easily integrated to 
\begin{equation}
\mu=\mu(0) e^{-\frac{1}{16 {\pi}^{2}} \int_{t}^{0}{\left(\sum_{i=1}^{3} (3|\lambda_{u_{i}}|^{2}+3|\lambda_{d_{i}}|^{2}+|\lambda_{\nu_{i}}|^{2}+|\lambda_{e_{i}}|^{2}) -\frac{3}{5}g_{1}^{2}-3g_{2}^{2}\right) }} \ .
\label{103}
\end{equation}
As discussed earlier, a constant field-independent $\mu(0)$ parameter cannot arise in a supersymmetric string vacuum since the Higgs fields are zero modes.  However, a $\mu$-term can arise from higher-dimensional couplings to moduli. Although typically chosen to be of electroweak order, in this paper, to emphasize the $B$-$L$/electroweak hierarchy and simplify the calculation, we will choose $\mu(0)$ to be substantially smaller, making its effects sub-dominant.  It is clear from  (\ref{103}) that $\mu$ runs slowly over the scaling range and, thus, once chosen to be sub-dominant at the unification scale it remains so throughout the scaling range.  The exact meaning of the term ``sub-dominant'' will be given in Appendix C.

\subsection*{The $B$ Parameter:}

The $B$ parameter enters the potential for the Higgs scalars $H$, $\bar{H}$
through the soft supersymmetry breaking term $V_{2s}$ in (\ref{12}). The RGE for $B$ is given by \cite{Yb}
\begin{eqnarray}
16 {\pi}^{2} \frac{dB}{dt}&=&  B \left(\sum_{i=1}^{3} (3|\lambda_{u_{i}}|^{2}+3|\lambda_{d_{i}}|^{2}+|\lambda_{\nu_{i}}|^{2}+|\lambda_{e_{i}}|^{2}) -\frac{3}{5}g_{1}^{2}-3g_{2}^{2}\right)\nonumber \\ 
&+&  \mu \Bigg( \sum_{i=1}^{3} ( 6 A_{u_{i}} \lambda^{*}_{u_{i}} +  A_{d_{i}} \lambda^{*}_{d_{i}} + 2 A_{e_{i}} \lambda^{*}_{e_{i}} + 2 A_{\nu_{i}} \lambda^{*}_{\nu_{i}} )  \label{132} \\  
&  + & 6 g_{2}^{2} M_{2} + \frac{6}{5} g_{1}^{2} M_{1} \Bigg)   \ . \nonumber
\end{eqnarray}
Recalling the discussion in the previous subsection, we take the term proportional to $\mu$ to be sub-leading.  Equation (\ref{132}) is then solved by 
\begin{equation}
B=B(0) e^{-\frac{1}{16 {\pi}^{2}} \int_{t}^{0}{\left(\sum_{i=1}^{3} (3|\lambda_{u_{i}}|^{2}+3|\lambda_{d_{i}}|^{2}+|\lambda_{\nu_{i}}|^{2}+|\lambda_{e_{i}}|^{2}) -\frac{3}{5}g_{1}^{2}-3g_{2}^{2}\right) }} \ .
\label{133}
\end{equation}
Here, unlike the constant, field-independent $\mu(0)$ parameter, the dimension two $B(0)$ parameter arises from supersymmetry breaking and need not vanish. Using (\ref{34}), (\ref{109}) and (\ref{110}), the RGE for $B$ can be approximated by
\begin{equation}
B \simeq B(0)e^{ -\frac{3}{16 {\pi}^{2}} \int_{t}^{0}{ |\lambda_{u_{3}}|^{2}}} \ .
\label{134}
\end{equation}
This parameter will make its appearance in the analysis of the Higgs potential below.
The assumption that the term proportional to $\mu$ is sub-leading is easily checked using (\ref{128})
and the relative sizes of $B$ and $\mu$ presented in Appendix C.

\subsection*{ Higgs and Squark Masses at the $B$-$L$ Breaking Vacuum:}

At the end of the previous section we showed that, when restricted to slepton scalars only, the potential energy has a local minimum given in (\ref{99}) and (\ref{100}). Clearly, however, to understand the stability of this minimum it is essential to extend this analysis to the entire field space; that is, to include all Higgs  and squark scalars and  as well as the sleptons. Given the RGE solutions in the previous subsections, one can now study the full scalar potential for any scale in the range (\ref{20}).

Let us begin by considering the quadratic mass terms, not at the origin of field space but, rather, at the slepton minimum given in (\ref{99}) and (\ref{100}). The relevant part of the scalar potential is still (\ref{73}). However, this is now evaluated for all scalar fields near the slepton VEVs. Note that, in principle, the $|\lambda_{\nu_{3}}|^{2}|H|^{2}|\nu_{3}|^{2}$ term in the $|\partial_{L_{3}}W|^{2}$ contribution to the potential, where $W$ is given in (\ref{6}), could play a role in selecting the vacuum. However, since $\lambda_{\nu_{3}} \ll g_{4}$, this term can safely be ignored. Similarly, under the assumptions (\ref{32}) and (\ref{33}), one can neglect the $A_{\nu_{3}}L_{3}H \nu_{3} +hc$ term in $V_{3s}$. Expanding out $V$ using the $B$-$L$ quantum numbers listed in (\ref{4}), the quadratic terms at any scale $t$ are
\begin{equation}
V= \dots + V_{m_{\rm slepton}^{2}}+ V_{m_{\rm Higgs}^{2}}+ V_{m_{\rm squarks}^{2}}+ \dots \ ,
\label{141}
\end{equation}
where $V_{m_{\rm slepton}^{2}}$ is given in (\ref{96}), (\ref{97}). The Higgs contribution to the quadratic potential is
\begin{equation}
V_{m_{\rm Higgs}^{2}}= m_{H}^{2} |H|^{2}+ m_{\bar{H}}^{2} |\bar{H}|^{2}-B(H{\bar{H}}+hc)
\label{145}
\end{equation}
where $m_{H}^{2}$, $m_{\bar{H}}^{2}$ and $B$ are given in (\ref{129}), (\ref{140}) and (\ref{134}) respectively. Note from (\ref{129}), (\ref{140}) that for $t \ll 0$ the quantity 
$|m_{\bar{H}}^{2}-m_{H}^{2}|$ becomes non-zero and large. Henceforth, we will assume that in this range of $t$ the coefficient $B$ is such that
\begin{equation}
4\left( \frac{B}{m_{\bar{H}}^{2}-m_{H}^{2}}\right)^{2} \ll 1\ .
\label{146}
\end{equation}
This is easily arranged by adjusting $B(0)$. For $t \ll 0$, the Higgs mass matrix (\ref{145}) can be diagonalized to
\begin{equation}
V_{m_{\rm Higgs}^{2}}= m_{H'}^{2} |H'|^{2}+ m_{{\bar{H}}'}^{2} |{\bar{H}}'|^{2} \ , 
\label{147}
\end{equation}
where
\begin{equation}
m_{H'}^{2} \simeq m_{H}^{2}-m_{\bar{H}}^{2}\left( \frac{B}{m_{\bar{H}}^{2}-m_{H}^{2}}\right)^{2}, \quad
m_{{\bar{H}}'}^{2} \simeq m_{\bar{H}}^{2}-m_{H}^{2}\left( \frac{B}{m_{\bar{H}}^{2}-m_{H}^{2}}\right)^{2}
\label{148}
\end{equation}
and
\begin{equation}
H' \simeq H + \left( \frac{-B}{m_{\bar{H}}^{2}-m_{H}^{2}}\right) {\bar{H}}^*, \quad 
{\bar{H}}' \simeq  \left( \frac{B}{m_{\bar{H}}^{2}-m_{H}^{2}}\right)H^* +{\bar{H}} \ .
\label{149}
\end{equation}
Here, ${{\bar{H}}}^{*A}=g^{A{\dot{C}}} \epsilon_{{\dot{C}}{\dot{D}}}{\bar{H}}^{*{\dot{D}}}$ and
${H}^{*A}=g^{A{\dot{C}}} \epsilon_{{\dot{C}}{\dot{D}}}{H}^{*{\dot{D}}}$ have the same $SU(2)_{L} \times
U(1)_{Y}$ transformations as $H$ and $\bar{H}$ respectively. In component fields
\begin{equation}
{{\bar{H}}}^{*}=({\bar{H}}^{-*}, -{\bar{H}}^{0*}), \quad H^{*}=(H^{0*}, -H^{+*}) \ .
\label{cat1}
\end{equation}

It follows from 
(\ref{129}), (\ref{140}), (\ref{146}) and (\ref{148}) that anywhere in the range $t \ll 0$, and specifically at $t_{B-L}$,
\begin{equation}
m_{{\bar{H}}'}^{2} \simeq m_{\bar{H}}^{2}=m_{H}(0)^{2} > 0 \ .
\label{150}
\end{equation}
Importantly, however, we see from (\ref{129}), (\ref{140}) that as $t$ becomes more negative $m_{H}^{2}$ can approach, become equal to and finally become smaller than $m_{\bar{H}}^{2}\left( B/ (m_{\bar{H}}^{2}-m_{H}^{2})\right)^{2}$.  This corresponds to $m_{H'}^{2}$ being positive, zero and negative respectively. The vanishing of $m_{H'}^{2}$ signals the onset of electroweak symmetry breaking.
Clearly, to evaluate $m_{H'}^{2}$ at $t_{B-L}$ we must solve equation (\ref{123A}) or, equivalently, (\ref{129}) for the running of $m_{H}^{2}$. The solution of this equation will depend on two arbitrary parameters, $m_{H}(0)^{2}$ and, using (\ref{24}), $M_{3}(0)$. It follows from the exponential form of (\ref{129}) and the fact that integral (\ref{120}) is non-negative that increasing $|M_{3}(0)|$ decreases $m_{H}(t)^{2}$ for any fixed value of $t$, and vice versa. Hence, specifying $|M_{3}(0)|$ is equivalent to specifying the value of 
$m_{H}(t)^{2}$ at some fixed $t$. For reasons discussed in in detail below, the physics is most transparent if we choose this to be the electroweak scale corresponding to $t_{EW}\simeq -33.3$.  Specifically, we take
\begin{equation}
m_{H}(t_{EW})^{2}=\frac{m_{H}(0)^{2}}{{\cal{T}}'^{2}},  
\label{buddy2}
\end{equation}
where
\begin{equation}
{\cal{T}}'^{2}= \frac{{\cal{T}}^{2}}{1-\Delta^{2}} \ .
\label{home1}
\end{equation}
${\cal{T}}$ is defined by
\begin{equation}
{\cal{T}}^{2} \equiv \left( \frac{B}{m_{\bar{H}}^{2}-m_{H}^{2}}\right)^{-2} \stackrel{>}{\sim} 40 \ ,
\label{buddy3}
\end{equation}
thus explicitly satisfying constraint (\ref{146}). $\Delta$ is a parameter with the range $0 < {\Delta}^{2} < 1$. The upper bound follows from our requirement that $m_{H}^{2}$ be positive over the entire physical scaling range and, hence, positive at $t_{EW}$. Note from (\ref{148}), (\ref{150}),  (\ref{buddy2}), (\ref{home1}) and (\ref{buddy3}) that
\begin{equation}
m_{H'}(t_{EW})^{2}=\frac{-\Delta^{2}}{{\cal{T}}^{2}} \ m_{H}(0)^{2} \ .
\label{home2}
\end{equation}
It follows that electroweak symmetry will be broken at $t_{EW}$ only if $\Delta^{2}$ is strictly positive. Hence, the lower bound on this parameter.

 The role of (\ref{buddy2}) in electroweak breaking will be thoroughly discussed in the next subsections. Its relationship to $|M_{3}(0)|$ and, hence, to the solution of (\ref{123A}) for $m_{H}(t)^{2}$ will be derived in Appendix A. Here, we will simply state the result that specifying this value at $t_{EW}$ is equivalent to choosing
\begin{equation}
|M_{3}(0)|^{2}=.0352(1-\frac{11.5}{{\cal{T}'}^{2}})m_{H}(0)^{2} \ .
\label{cat1}
\end{equation}
Using this expression for $|M_{3}(0)|$, we can solve (\ref{123A}) numerically for any fixed value of ${\cal{T}'}$. The numerical results can then be fit to a smooth curve. We find that the numerical data is well-represented over the entire scaling regime by
\begin{equation}
m_{H}(t)^{2} = \big( 1 -(1 - \frac{1}{{\cal{T}'}^{2}} )( \frac{t_{EW}-b}{t_{EW}} ) ( \frac{t}{t-b} )\big)m_{H}(0)^{2}  \ ,
\label{funfit}
\end{equation}
where b is a function of ${\cal{T'}}$ of the form
\begin{equation}
b({\cal{T'}}) =19.9(1-\frac{.186}{{\cal{T}'}-3.69}) \ .
\label{funfit2}
\end{equation}
Note that at $t=0$ and $t=t_{EW}$, $m_{H}^{2}$ is given by $m_{H}(0)^{2}$ and $m_{H}(0)^{2}/{\cal{T}'}^{2}$ respectively, as it must be. One can now use expression (\ref{funfit}) to evaluate $m_{H}^{2}$ and, hence, $m_{H'}^{2}$ at $t_{B-L}$ for any ${\cal{T}}$ and parameter $\Delta^{2}$. It suffices here to give the lower bound. To do this, first note that for $\Delta^{2}=1$ the ${\cal{T}}$ dependence drops out of (\ref{funfit}) and (\ref{funfit2}). Furthermore, it follows from (\ref{home2}) that this corresponds to the largest negative $H'^{2}$ squared mass at $t_{EW}$ and, hence, a ${\cal{T}}$-dependent lower bound on $m_{H'}^{2}$ at $t_{B-L}$. Taking $\Delta^{2}=1$, we find
\begin{equation}
m_{H}(t_{B-L})^{2}= .0565 \ m_{H}(0)^{2}
\label{home3}
\end{equation}
and, hence,
\begin{equation}
m_{H'}(t_{B-L})^{2}= .0565( 1-\frac{17.7}{{\cal{T}}^{2}}) \  m_{H}(0)^{2} \ .
\label{home4}
\end{equation}
Since, from (\ref{buddy3}), ${\cal{T}}^{2} \stackrel{>}{\sim} 40$, this expression is always positive. We conclude that at $t_{B-L}$, for any ${\cal{T}}$ and $0< \Delta^{2} <1$, 
\begin{equation}
m_{H'}^{2} > .0565( 1-\frac{17.7}{{\cal{T}}^{2}}) \  m_{H}(0)^{2} >0 \ .
\label{home5}
\end{equation}
Finally, note that evaluated at the slepton vacuum (\ref{99}), (\ref{100}) the diagonalized quadratic Higgs potential remains
\begin{equation}
V_{m_{\rm Higgs}^{2}}= \langle m_{H'}^{2} \rangle |H'|^{2}+ \langle m_{{\bar{H}}'}^{2} \rangle 
|{\bar{H}}'|^{2} \ , 
\label{home6}
\end{equation}
with
\begin{equation}
 \langle m_{H'}^{2} \rangle = m_{H'}^{2}, \qquad \langle m_{{\bar{H}}'}^{2} \rangle= m_{{\bar{H}}'}^{2} \ .
\label{home7}
\end{equation}
It then follows from (\ref{home5}) and (\ref{150}) respectively that at $t_{B-L}$
\begin{equation}
 \langle m_{H'}^{2} \rangle > .0565( 1-\frac{17.7}{{\cal{T}}^{2}}) \  m_{H}(0)^{2} , \quad
\langle m_{{\bar{H}}'}^{2} \rangle = m_{H}(0)^{2} 
\label{home8}
\end{equation}
and that they are both positive.

The squark contribution to the quadratic potential is
\begin{equation}
V_{m_{\rm squark}^{2}}=\sum_{i=1}^{3} (\langle m_{Q_{i}}^{2} \rangle |Q_{i}|^{2}+\langle m_{u_{i}}^{2} \rangle |u_{i}|^{2}+\langle m_{d_{i}}^{2} \rangle |d_{i}|^{2}) \ ,
\label{142}
\end{equation}
with
\begin{equation}
\langle m_{Q_{i}}^{2} \rangle = m_{Q_{i}}^{2}+\frac{1}{4} g_{4}^{2} \langle \nu_{3} \rangle^{2}
\label{143}
\end{equation}
and
\begin{eqnarray}
\langle m_{u_{i}}^{2} \rangle &= & m_{u_{i}}^{2}-\frac{1}{4} g_{4}^{2} \langle \nu_{3} \rangle^{2}, \nonumber \\
\langle m_{d_{i}}^{2} \rangle & =& m_{d_{i}}^{2}-\frac{1}{4} g_{4}^{2} \langle \nu_{3} \rangle^{2} \ .
 \label{144}
\end{eqnarray}
Here $m_{Q_{i}}^{2}$ are given in (\ref{118A}), (\ref{130}), $m_{u_{i}}^{2}$ are given in (\ref{119A}), (\ref{131}) and $m_{d_{i}}^{2}$ are given in (\ref{138}).

Using this result, one can now evaluate the squark masses in (\ref{143}) and (\ref{144}). Let us begin by  computing the second term on the right hand side of these equations. Using (\ref{16}), (\ref{21}), (\ref{93}) and (\ref{95}), we find that at $t_{B-L}$
\begin{equation}
\frac{1}{4}  g_{4}^{2} \langle \nu_{3} \rangle^{2}= \ \frac{4}{3} \ m_{\nu}(0)^{2} \ .
\label{166}
\end{equation}
First consider $\langle m_{Q_{1,2}}^{2} \rangle$. Recall from (\ref{130}) that
\begin{equation}
m_{Q_{1,2}}^{2}  \simeq -\frac{2}{3 {\pi}^{2}} \int_{0}^{t}{g_{3}^{2}|M_{3}|^{2}}
+\frac{1}{64 {\pi}^{2}} \int_{0}^{t}{g_{4}^{2}{\cal{S}'}_{1}} +\frac{m_{H}(0)^{2}}{2} \ . 
\label{130B}
\end{equation}
For any value of $t$, the first term can be evaluated using (\ref{120}), (\ref{cat1}) and the second with  (\ref{93}), (\ref{121}). We find that at $t_{B-L}$
\begin{eqnarray}
& & -\frac{2}{3 {\pi}^{2}} \int_{0}^{t_{B-L}}{g_{3}^{2}|M_{3}|^{2}}= .143 \big( 1 - \frac{11.5}{{\cal{T'}}^{2}}  \big) m_{H}(0)^{2} , \label{me1} \\
& &  \qquad \ \frac{1}{64 {\pi}^{2}} \int_{0}^{t_{B-L}}g_{4}^{2}{\cal{S}'}_{1}=  - 1.66\  m_{\nu}(0)^{2}  \label{me2}
\end{eqnarray}
and, hence, 
\begin{equation}
m_{Q_{1,2}}^{2}  \simeq .643(1 - \frac{2.57}{{\cal{T'}}^{2}}) m_{H}(0)^{2} - 1.66 \ m_{\nu}(0)^{2} \ .
\label{me3}
\end{equation}
Using this and (\ref{166}) in (\ref{143}) gives
\begin{equation}
\langle m_{Q_{1,2}}^{2} \rangle= .643(1 - \frac{2.57}{{\cal{T'}}^{2}}) m_{H}(0)^{2} - .336 \ m_{\nu}(0)^{2} \ .
\label{me4}
\end{equation}
To further evaluate this squared mass and to determine whether or not it is positive, one must give an explicit relationship between the parameters $m_{\nu}(0)^{2}$ and $m_{H}(0)^{2}$. This will be discussed in detail in Appendix B. Here, we will simply use the result. By demanding that all squark mass squares remain minimally positive at any scale $t_{EW} \leq t \leq 0$, we find that
\begin{equation}
m_{\nu}(0)^{2}=.864(1- \frac{2.25}{{\cal{T}'}^{2}} )m_{H}(0)^{2} \ .
\label{me5}
\end{equation}
Using this to eliminate the $m_{H}(0)^{2}$ parameter, expression (\ref{me4}) becomes
\begin{equation}
\langle m_{Q_{1,2}}^{2} \rangle= .408(1 -  \frac{.583}{{\cal{T'}}^{2}} ) m_{\nu}(0)^{2} \ .
\label{me6}
\end{equation}
Note that this expression is positive for all parameters ${\cal{T}}^{2} \stackrel{>}{\sim} 40$ and $0<\Delta^{2}<1$. Also, for sake of comparison to the sneutrino VEV, we will write all masses at $t_{B-L}$ in terms of $m_{\nu}(0)^{2}$.  The form of $m_{u_{1,2}}^{2}$ and $m_{d_{i}}^{2}$ for $i=1,2,3$ given in (\ref{131}) and (\ref{138}) respectively are similar to that of $m_{Q_{1,2}}^{2}$ and can be evaluated using (\ref{me1}), (\ref{me2}) and (\ref{me5}). Using these results and (\ref{166}), we find from (\ref{144}) that 
\begin{equation}
\langle m_{u_{1,2}}^{2} \rangle = \langle m_{d_{i}}^{2} \rangle = 
1.08(1 - \frac{0.223}{{\cal{T'}}^{2}} ) m_{\nu}(0)^{2}  \ .
\label{me7}
\end{equation}
These mass squares are positive for all ${\cal{T}}^{2} \stackrel{>}{\sim} 40$ and $0<\Delta^{2}<1$.

Now consider $\langle m_{Q_{3}}^{2} \rangle$. Recall from (\ref{118A}) that
\begin{equation}
m_{Q_{3}}^{2}  \simeq  \frac{1}{3}m_{H}^{2}-\frac{2}{3 {\pi}^{2}} \int_{0}^{t}{g_{3}^{2}|M_{3}|^{2}}+\frac{1}{64 {\pi}^{2}} \int_{0}^{t}g_{4}^{2}{\cal{S}'}_{1}+\frac{1}{6}m_{H}(0)^{2} \ . 
\label{118B} 
\end{equation}
The second and third terms are given in (\ref{me1}) and (\ref{me2}) respectively. Unlike the masses evaluated above, however, $ m_{Q_{3}}^{2} $ also depends on the mass parameter $m_{H}^{2}$. This can be evaluated using (\ref{funfit}), (\ref{funfit2}) for any value of $t$. At $t_{B-L}$ we find 
\begin{equation}
m_{H}^{2} \simeq .0565(1-\frac{0.101}{{\cal{T}'}}+\frac{16.3}{{\cal{T}'}^{2}}) m_{H}(0)^{2} \ .
\label{try1}
\end{equation}
Using this together with (\ref{me1}) and  (\ref{me2}), 
$m_{Q_{3}}^{2}$ at $t=t_{B-L}$ is given by 
\begin{equation}
m_{Q_{3}}^{2} \simeq  -1.66 m_{\nu}(0)^{2} + 0.328 (1-\frac{0.00629}{{\cal{T}'}}- \frac{4.09}{{\cal{T}'}^{2}})  m_{H}(0)^{2} \ , 
\label{167}
\end{equation}
which, using (\ref{me5}), becomes
\begin{equation}
m_{Q_{3}}^{2} \simeq -1.29 ( 1 +  \frac{0.00185}{{\cal{T}'}}+ \frac{0.545}{{\cal{T}'}^{2}}) m_{\nu}(0)^{2} \ .
\label{168}
\end{equation}
Similarly, $m_{u_{3}}^{2}$ given by (\ref{119A}) is found to be
\begin{equation}
m_{u_{3}}^{2} \simeq 1.68 ( 1 -  \frac{0.00284}{{\cal{T}'}}- \frac{0.691}{{\cal{T}'}^{2}}) m_{\nu}(0)^{2} 
\label{170}
\end{equation}
at $t=t_{B-L}$. Using these results and (\ref{166}), we find from (\ref{143}),(\ref{144}) that 
\begin{eqnarray}
\langle m_{Q_{3}}^{2} \rangle &\simeq &0.0435 ( 1 -  \frac{0.0549}{{\cal{T}'}}- \frac{16.2}{{\cal{T}'}^{2}}) m_{\nu}(0)^{2}, \nonumber \\  \langle m_{u_{3}}^{2}\rangle &\simeq & 0.353 ( 1 -  \frac{0.0136}{{\cal{T}'}}- \frac{3.30}{{\cal{T}'}^{2}}) m_{\nu}(0)^{2}  \ .
\label{172}
\end{eqnarray}
Note that they are both  positive for all ${\cal{T}}^{2} \stackrel{>}{\sim} 40$ and $0<\Delta^{2}<1$. Putting everything together, the squark masses, expressed in terms of $m_{\nu}(0)^{2}$, are given in (\ref{me6}), (\ref{me7}) and (\ref{172}) and are all positive.

From (\ref{98}), (\ref{150}),(\ref{home5}), (\ref{me6}), (\ref{me7}) and (\ref{172}) we conclude that at $t=t_{B-L}$ the vacuum specified by
\begin{equation}
\langle \nu_{1,2} \rangle = 0, \quad \langle \nu_{3} \rangle =\frac{2 \ m_{\nu}(0)}{\sqrt{\frac{3}{4}}g_{4}}, \quad
\langle e_{i} \rangle=\langle L_{i} \rangle = 0
\label{174}
\end{equation}
\begin{equation}
\langle Q_{i} \rangle = \langle u_{i} \rangle = \langle d_{i} \rangle = 0
\label{175}
\end{equation} 
and
\begin{equation}
\langle H' \rangle = \langle {\bar{H}'} \rangle = 0
\label{176}
\end{equation}
is a local minimum of the potential energy. It has positive mass squares in every field direction
including $H'$ and $\bar{H}'$, signaling that, at $t=t_{B-L}$, electroweak symmetry has not yet been spontaneously broken.

\subsection*{Electroweak Symmetry Breaking:}

To explore the breaking of electroweak symmetry, one must add to $V_{m_{Higgs}^{2}}$ in (\ref{145}) all other relevant interactions involving the Higgs fields. It follows from previous discussions that 
the relevant part of the Higgs potential is
\begin{equation}
V=V_{m_{Higgs}^{2}}+\frac{1}{2}D_{Y}^{2}+\frac{1}{2} \sum_{a=1}^{3} D_{SU(2)_{L}a}^{2} 
\label{177}
\end{equation}
where, written in the mass diagonal fields $H'$ and ${\bar{H}}'$ defined in (\ref{149}), 
\begin{equation}
D_{Y}=\sqrt{\frac{3}{5}} \frac{g_{1}}{2}(|H'|^{2}-|{\bar{H}}'|^{2}) \ ,
\label{178}
\end{equation}
\begin{equation}
D_{SU(2)_{L}a}=\frac{g_{2}}{2} ( {H'}^{\dagger} \sigma_{a}H'+{{\bar{H}}}^{ \prime \dagger} \sigma_{a}{\bar{H}}') 
\label{179}
\end{equation}
and we have dropped terms of ${\cal{O}}({\cal{T}}^{-1})$. The parameters in this potential should now be evaluated, not at $t_{B-L}$, but, rather, at the electroweak scale $M_{EW} \simeq 10^{2}GeV$ which corresponds to a $t_{EW} \simeq -33.3$. The gauge couplings are easily evaluted there using (\ref{18})
and (\ref{19}) and found to be
\begin{equation}
\frac{g_{1}^{2}(t_{EW})}{g(0)^{2}}= .405, \quad \frac{g_{2}^{2}(t_{EW})}{g(0)^{2}}= .818 \ .
\label{180}
\end{equation}
Also, it follows from (\ref{140}) and (\ref{home2}) that
\begin{equation}
m _{{\bar{H}}'}(t_{EW})^{2} \simeq m_{H}(0)^{2}, \quad m_{H'}(t_{EW})^{2}= \frac{-\Delta^{2}}{{\cal{T}}^{2}}m_{H}(0)^{2} 
\label{181}
\end{equation}
where ${\cal{T}}^{2} \stackrel{>}{\sim} 40$ and $0<\Delta^{2}<1$.

We can now proceed to minimize potential (\ref{177}). Writing
\begin{equation}
H'=(H'^{+}, H'^{0}), \quad {\bar{H}}' =( {\bar{H}}'^{0},  {\bar{H}}'^{-}) \ ,
\label{189}
\end{equation}
potential (\ref{177}) becomes
\begin{eqnarray}
V&=&  - \frac{\Delta^{2}}{{\cal{T}}^{2}}m_{H}(0)^{2}(|H'^{0}|^{2}+|H'^{+}|^{2})+ m_{H}(0)^{2} (|{\bar{H}}'^{0}|^{2}+|{\bar{H}}'^{-}|^{2}) \nonumber \\
& + &\frac{1}{8}(\frac{3}{5})g_{1}^{2}(|H'^{0}|^{2}+|H'^{+}|^{2} - |{\bar{H}}'^{0}|^{2}- |{\bar{H}}'^{-}|^{2})^{2} \label{190} \\
& + & \frac{1}{8}g_{2}^{2}([|H'^{0}|^{2}+|H'^{+}|^{2} - |{\bar{H}}'^{0}|^{2}-|{\bar{H}}'^{-}|^{2}]^{2}  \nonumber \\
& + &4|H'^{+} {\bar{H}}'^{0*}+ H'^{0}{\bar{H}}'^{-*}|^{2})  \nonumber \ . 
\label{190}
\end{eqnarray}
This is easily minimized to give
\begin{equation}
{\langle \langle} H'^{0} \rangle \rangle =\frac{2\Delta m_{H}(0)}{{\cal{T}}\sqrt{\frac{3}{5}g_{1}(t_{EW})^{2}+g_{2}(t_{EW})^{2}}}, \quad \langle \langle H'^{+} \rangle \rangle = 0 \ ,
\label{191}
\end{equation}
and
\begin{equation}
\langle \langle {\bar{H}}'^{0} \rangle \rangle =\langle \langle {\bar{H}}'^{-} \rangle \rangle = 0 \ ,
\label{192}
\end{equation}
where the double bracket $\langle \langle \ \rangle \rangle$ indicates the vacuum at $t_{EW}$. 
It is straightforward to compute the squared masses of the radial fluctuation $\delta H'^{0}$ and complex ${\bar{H}}'^{0}$, ${\bar{H}}'^{-}$ fields at this vacuum. We find that
\begin{equation}
V=\langle \langle m_{H'^{0}}^{2} \rangle \rangle |\delta H'^{0}|^{2} +\langle \langle m_{{\bar{H}}'^{0}}^{2}\rangle \rangle |{\bar{H}}'^{0}|^{2} +\langle \langle m_{{\bar{H}}'^{-}}^{2} \rangle \rangle |{ \bar{H}}'^{-}|^{2} +\dots
\label{193}
\end{equation}
where
\begin{equation}
\langle \langle m_{H'^{0}}^{2} \rangle \rangle= 4\frac{\Delta^{2}m_{H}(0)^{2}}{{\cal{T}}^{2}}
\label{194}
\end{equation}
and
\begin{equation}
\langle \langle m_{{\bar{H}}'^{0}}^{2} \rangle \rangle=(1-\frac{\Delta^{2}}{{\cal{T}}^{2}})m_{H}(0)^{2}, \quad \langle \langle m_{{\bar{H}}'^{-}}^{2} \rangle \rangle=(1-\frac{\Delta^{2}}{{\cal{T}}^{2}}(\frac{g_{1}^{2}-g_{2}^{2}}{g_{1}^{2}+g_{2}^{2}}))m_{H}(0)^{2} \ .
\label{195}
\end{equation}
Note that the Higgs mass squares are positive for all ${\cal{T}}^{2} \stackrel{>}{\sim} 40$ and $0<\Delta^{2}<1$. Evaluated at this minimum, the phase of $H'^{0}$ and $H'^{-}$, $H'^{-*}$ are Goldstone bosons which are ``eaten'' by the Higgs mechanism to give mass to the $Z$ and $W^{\pm}$ vector bosons. For example, the $Z$ mass is given by
\begin{equation}
M_Z=\frac{\sqrt{2} \Delta m_{H}(0)}{{\cal{T}}} \simeq 91GeV \ .
\label{train1}
\end{equation}

Although the mass eigenstate basis $H'$, ${\bar{H}}'$ is the most natural for analyzing this vacuum, it is of some interest to express it in terms of the original 
$H$ and $\bar{H}'$ fields. Using (\ref{5}), (\ref{149}) and (\ref{buddy3}), we find 
\begin{equation}
\langle \langle H^{+} \rangle \rangle = \langle \langle {\bar{H}}^{-} \rangle \rangle = 0 
\label{197}
\end{equation}
and, to leading order, that
\begin{equation}
\langle \langle H^{0} \rangle \rangle =\frac{2\Delta m_{H}(0)}{{\cal{T}}\sqrt{\frac{3}{5}g_{1}(t_{EW})^{2}+g_{2}(t_{EW})^{2}}}, \quad \langle \langle {\bar{H}}^{0} \rangle \rangle = \frac{1}{{\cal{T}}} \langle \langle H^{0} \rangle \rangle \ .
\label{198}
\end{equation}
Note that the condition $\langle \langle {\bar{H}}'^{0} \rangle \rangle =0$ in (\ref{192}) does {\it not} imply the vanishing of $ \langle \langle {\bar{H}}^{0} \rangle \rangle$. Rather, $\langle \langle {\bar{H}}^{0} \rangle \rangle$ is non-zero and related to $\langle \langle H^{0} \rangle \rangle$ through the ratio
 \begin{equation}
 \frac{ \langle \langle H^{0} \rangle \rangle}{\langle \langle {\bar{H}}^{0} \rangle \rangle }\equiv tan\beta
={\cal{T}} +{\cal{O}}({\cal{T}}^{-1}) \ .
\label{199}
\end{equation}
We have indicated the ${\cal{O}}({\cal{T}}^{-1})$  contribution to emphasize that although $tan\beta={\cal{T}}$ {\it to leading order}, this relationship breaks down at higher order in ${\cal{T}}^{-1}$. We refer the reader to Appendix C for a more detailed discussion of this point.

We conclude that electroweak symmetry is spontaneously broken at scale $t_{EW}$ by the non-vanishing $H'^{0}$ vacuum expectation value in (\ref{191}). This vacuum has a non-vanishing value of $tan\beta$ which, using the assumption for ${\cal{T}}^{2}$ given in (\ref{buddy3}), satisfies
\begin{equation}
tan\beta \stackrel{>}{\sim} 6.32 \ .
\label{201}
\end{equation}
As far as the Higgs fields are concerned, the vacuum specified in (\ref{191}) and (\ref{192})
is a stable local minimum. 
As a check on our result, choose $\mu^{2}$ of order ${\cal{O}}({\cal{T}}^{-4})$ or smaller, that is, non-vanishing but sub-dominant in all equations.  Then (\ref{150}),(\ref{buddy2}),(\ref{home1}) and (\ref{buddy3}) satisfy the constraint equations, given, for example, in \cite{Zc}, for the Higgs potential to be bounded below and have a negative squared mass at the origin. Furthermore, to the order in ${\cal{T}}^{-1}$ we are working, (\ref{train1}) and (\ref{198}) for the Higgs vacuum satisfy the minimization conditions in \cite{Zc}.
This is shown in detail in Appendix C.

\subsection*{Slepton and Squark Masses at the EW Breaking Vacuum:}

Clearly, to understand the complete stability of this minimum, it is essential to extend this analysis to the entire field space; that is, to include all slepton and squark scalars as well as the Higgs fields. The relevant part of the potential energy is
\begin{equation}
V=V_{2s}+\frac{1}{2}D_{B-L}^{2}+\frac{1}{2}D_{Y}^{2}+\frac{1}{2} \sum_{a=1}^{3} D_{SU(2)_{L}a}^{2} \ , 
\label{202}
\end{equation}
where $V_{2s}$, $D_{B-L}$, $D_{Y}$ are given in (\ref{12}), (\ref{10}), (\ref{9}) respectively and 
$D_{SU(2)_{L}a}$ is the extension of (\ref{179}) to include all slepton and squark doublets. Note that, in principle, the superpotential (\ref{6}) can also contribute to the squark/slepton potential energy. However, it follows from (\ref{34}) and (\ref{109}),(\ref{110}) that such terms will be negligibly small compared to those in (\ref{202}) {\it with the notable exception of the third family up-squarks}. We will take this contribution into account when it arises below.
Now expand $V$ to quadratic order in the fields using the quantum numbers listed in (\ref{2}), (\ref{3}) and (\ref{4}). We find that, evaluated at the slepton and Higgs VEVs in (\ref{99}), (\ref{100}) and (\ref{191}), (\ref{192}) respectively, 
\begin{equation}
V=\dots+V_{m_{Higgs}^{2}} +V_{m_{sleptons}^{2}}+V_{m_{squarks}^{2}} +\dots \ ,
\label{203}
\end{equation}
where $V_{m_{Higgs}^{2}} $ is given in (\ref{193}), (\ref{194}) and (\ref{195}).

The slepton contribution to the quadratic potential is
\begin{eqnarray}
V_{m_{sleptons}^{2}} & = & \langle \langle m_{\nu_{3}}^{2} \rangle \rangle |\delta \nu_{3}|^{2} +\langle \langle m_{\nu_{1,2}}^{2} \rangle \rangle |\nu_{1,2}|^{2}
+\sum_{i=1}^{3} (\langle \langle m_{e_{i}}^{2} \rangle \rangle|e_{i}|^{2} \nonumber \\
& & +\langle \langle m_{N_{i}}^{2} \rangle \rangle |N_{i}|^{2}+\langle \langle m_{E_{i}}^{2} \rangle \rangle |E_{i}|^{2}) + \dots \ ,
\label{204}
\end{eqnarray}
where
\begin{eqnarray}
\langle \langle m_{\nu_{3}}^{2} \rangle \rangle & = & \langle m_{\nu_{3}}^{2} \rangle, \ \langle \langle m_{\nu_{1,2}}^{2} \rangle \rangle= \langle m_{\nu_{1,2}}^{2} \rangle, \nonumber \\ 
\langle \langle m_{e_{i}}^{2} \rangle \rangle &= & \langle m_{e_{i}}^{2} \rangle+ \frac{1}{2}(\frac{3}{5})g_{1}^{2}\langle \langle H'^{0} \rangle \rangle^{2}, 
\label{205} \\
 \langle \langle m_{N_{i}}^{2} \rangle \rangle  &= & \langle m_{L_{i}}^{2} \rangle-\frac{1}{4}(\frac{3}{5}g_{1}^{2}+g_{2}^{2})\langle \langle H'^{0} \rangle \rangle^{2},   \nonumber \\
 \langle \langle m_{E_{i}}^{2} \rangle \rangle & = & \langle m_{L_{i}}^{2} \rangle-\frac{1}{4}(\frac{3}{5}g_{1}^{2}-g_{2}^{2}) \langle \langle H'^{0} \rangle \rangle^{2} \ . \nonumber
\end{eqnarray}
The squared masses 
$ \langle m_{\nu_{3}}^{2} \rangle$, $\langle m_{\nu_{1,2}}^{2} \rangle$, $\langle m_{e_{i}}^{2} \rangle$ and $\langle m_{L_{i}}^{2} \rangle$  were defined in (\ref{97}) and evaluated at $t_{B-L}$ in (\ref{98}). Now, however, these values must be corrected by scaling down to $t_{EW}$. Using (\ref{16}) and (\ref{18}), the slepton masses defined in (\ref{53})-(\ref{55}) can be evaluated at $t_{EW}$. Note that the parameters $A$, $C$ and, hence, ${\cal{S}'}_{1}(0)$ given in (\ref{88}),(\ref{90}) and (\ref{93}) remain the same. We find that
\begin{eqnarray}
& &m_{\nu_{3} \rm}^{2} = -4.38  m_{\nu}(0)^{2} , \quad m_{\nu_{1,2} \rm}^{2} = 77.8  m_{\nu}(0)^{2}, \label{air1} \\
& & \ m_{e_{i} \rm}^{2} =  0.625  m_{\nu}(0)^{2}, \quad m_{L_{i} \rm} = 11.4 m_{\nu}(0)^{2}. \nonumber
\end{eqnarray}
Using these values, slepton potential (\ref{94}) has a local minimum at
\begin{equation}
\langle \langle \nu_{3} \rangle \rangle = (1.05)\frac{2 m_{\nu}(0)}{\sqrt{\frac{3}{4}}g_{4}(t_{EW})}, \quad \langle \langle \nu_{1,2} \rangle \rangle= 0
\label{air2}
\end{equation}
\begin{equation}
\langle \langle e_{i} \rangle \rangle= \langle \langle L_{i} \rangle \rangle= 0 \quad i=1,2,3
\label{air4}
\end{equation}
We can now evaluate the squared slepton masses in (\ref{97}) at $t_{EW}$. We find 
\begin{eqnarray}
& &\langle m_{\nu_{3}}^{2} \rangle = 8.75 m_{\nu}(0)^{2} , \quad  
\langle m_{\nu_{1,2}}^{2} \rangle  =  82.2 m_{\nu}(0)^{2},
\nonumber \\
& &\  \langle m_{e_{i}}^{2} \rangle  =  5 m_{\nu}(0)^{2},  
\qquad  \quad \langle m_{L_{i}}^{2} \rangle  =  7 \ m_{\nu}(0)^{2}
\label{air5} \ .
\end{eqnarray}
Inserting these results into (\ref{205}) and using (\ref{191}) gives 
\begin{eqnarray}
\langle \langle m_{\nu_{3}}^{2} \rangle \rangle & = & 8.75   m_{\nu}(0)^{2}, \ \langle \langle m_{\nu_{1,2}}^{2} \rangle \rangle= 82.2   \ m_{\nu}(0)^{2} , \nonumber \\
 \langle \langle m_{e_{i}}^{2} \rangle \rangle & = & 5 m_{\nu}(0)^{2}+2\left(\frac{\frac{3}{5}g_{1}^{2}}{\frac{3}{5}g_{1}^{2}+g_{2}^{2}}\right)\frac{\Delta^{2}m_{H}(0)^{2}}{{\cal{T}}^{2}}, 
\label{206} \\
\langle \langle m_{N_{i}}^{2} \rangle \rangle  &= & 7 m_{\nu}(0)^{2}-\frac{\Delta^{2}m_{H}(0)^{2}}{{\cal{T}}^{2}},  \nonumber \\
\langle \langle m_{E_{i}}^{2} \rangle \rangle & = &  7 m_{\nu}(0)^{2}+\left( \frac{-\frac{3}{5}g_{1}^{2}+g_{2}^{2}}{\frac{3}{5}g_{1}^{2}+g_{2}^{2}} \right) \frac{\Delta^{2}m_{H}(0)^{2}}{{\cal{T}}^{2}} \ . \nonumber
\end{eqnarray}
Finally, using expression (\ref{me5}) which relates $m_{\nu}(0)^{2}$ to $m_{H}(0)^{2}$ and  (\ref{180}), these squared masses become
\begin{eqnarray}
\langle \langle m_{\nu_{3}}^{2} \rangle \rangle & = & 8.75  m_{\nu}(0)^{2}, \ \langle \langle m_{\nu_{1,2}}^{2} \rangle \rangle= 82.2  \ m_{\nu}(0)^{2} , \nonumber \\
 \langle \langle m_{e_{i}}^{2} \rangle \rangle & = & 5 m_{\nu}(0)^{2}(1+\Delta^{2}\frac{0.0791}{{\cal{T}}^{2}}) \simeq 5 m_{\nu}(0)^{2}, 
\label{207} \\
\langle \langle m_{N_{i}}^{2} \rangle \rangle  &= & 7 m_{\nu}(0)^{2}(1-\Delta^{2}\frac{0.123}{{\cal{T}}^{2}}) \simeq 7 m_{\nu}(0)^{2},  \nonumber \\
\langle \langle m_{E_{i}}^{2} \rangle \rangle & = &  7 m_{\nu}(0)^{2}(1+\Delta^{2}\frac{0.0669}{{\cal{T}}^{2}}) \simeq \ 7 m_{\nu}(0)^{2} \ . \nonumber
\end{eqnarray}
Note that the slepton squared masses  are positive for all ${\cal{T}}^{2} \stackrel{>}{\sim} 40$ and $0<\Delta^{2}<1$. 

Similarly, the squark contribution to the quadratic potential is
\begin{equation}
V_{m_{\rm squark}^{2}}=\sum_{i=1}^{3} (\langle \langle m_{U_{i}}^{2} \rangle \rangle |U_{i}|^{2}+\langle \langle m_{D_{i}}^{2} \rangle \rangle |D_{i}|^{2}+\langle \langle m_{u_{i}}^{2} \rangle \rangle |u_{i}|^{2}+\langle \langle m_{d_{i}}^{2} \rangle \rangle |d_{i}|^{2}) 
\label{208}
\end{equation}
with
\begin{eqnarray}
\langle\langle m_{U_{i}}^{2} \rangle\rangle & = & \langle m_{Q_{i}}^{2} \rangle+\left(\frac{\frac{1}{5}g_{1}^{2}-g_{2}^{2}}{4}+|\lambda_{u_{3}}|^{2}\delta_{i3} \right)\langle \langle H'^{0} \rangle \rangle^{2}, \nonumber \\
\langle \langle m_{D_{i}}^{2} \rangle\rangle & = & \langle m_{Q_{i}}^{2} \rangle+\left(\frac{\frac{1}{5}g_{1}^{2}+g_{2}^{2}}{4}\right)\langle \langle H'^{0} \rangle \rangle^{2}
\label{209}
\end{eqnarray}
and
\begin{eqnarray}
\langle \langle m_{u_{i}}^{2} \rangle \rangle &= & \langle m_{u_{i}}^{2} \rangle-\left( \frac{1}{5} g_{1
}^{2} -|\lambda_{u_{3}}|^{2}\delta_{i3} \right) \langle \langle H'^{0} \rangle \rangle^{2}, \nonumber \\
\langle \langle m_{d_{i}}^{2} \rangle \rangle & = & \langle m_{d_{i}}^{2} \rangle+\frac{1}{10}g_{1
}^{2}\langle \langle H'^{0} \rangle \rangle^{2} \ .
 \label{210}
\end{eqnarray}
Note that in the expressions for $\langle\langle m_{U_{3}}^{2} \rangle\rangle$ and $\langle \langle m_{u_{3}}^{2} \rangle \rangle$ we have included the non-negligible superpotential contribution. 
The squared masses $\langle m_{Q_{i}}^{2} \rangle$, $\langle m_{u_{i}}^{2} \rangle$ and $\langle m_{d_{i}}^{2} \rangle$ were defined in (\ref{143}), (\ref{144}) and evaluated at $t_{B-L}$ in (\ref{172}).
Now, however, these values must be corrected by scaling down to $t_{EW}$.  First, we must compute $g_{4}$ at $t_{EW}$.  Using (\ref{16}) and (\ref{19}) we find
\begin{equation}
\frac{g_{4}^{2}(t_{EW})}{g(0)^{2}}= .272 \ .
\label{g4tew}
\end{equation}
One can now evaluate the second term on the right hand side of (\ref{143}) and (\ref{144}) using (\ref{air2}). The result is
\begin{equation}
\frac{1}{4}  g_{4}^{2} \langle \nu_{3} \rangle ^{2}= 1.46 m_{\nu}(0)^{2} \ .
\label{war1}
\end{equation}
The squark masses defined in (\ref{118A}),(\ref{119A}),(\ref{130}),(\ref{131}) and (\ref{138}) can be evaluated at $t_{EW}$ using (\ref{116}),(\ref{117}),(\ref{114}),(\ref{115}),(\ref{123}), (\ref{138}) and (\ref{me5}). Adding these to (\ref{war1}), expressions (\ref{143}), (\ref{144}) become
\begin{eqnarray}
\langle m_{Q_{3}}^{2} \rangle & = & 0.132 ( 1-\frac{14.7}{{\cal{T'}}^{2}} ) m_{H}(0)^{2}, \nonumber \\
\langle m_{Q_{1,2}}^{2} \rangle & = & 0.465 ( 1-\frac{4.87}{{\cal{T'}}^{2}} ) m_{H}(0)^{2}, \nonumber \\
\langle m_{u_{3}}^{2} \rangle & = & 0.374 ( 1-\frac{7.72}{{\cal{T'}}^{2}} ) m_{H}(0)^{2}, \label{war2} \\
\langle m_{u_{1,2}}^{2} \rangle & = & 1.041 ( 1-\frac{3.42}{{\cal{T'}}^{2}} )  m_{H}(0)^{2}, \nonumber \\
\langle m_{d_{i}}^{2} \rangle & = & 1.041 ( 1-\frac{3.42}{{\cal{T'}}^{2}} )  m_{H}(0)^{2}. \nonumber
\end{eqnarray}
Inserting (\ref{war2}) and (\ref{191}) into (\ref{209}) and (\ref{210}), we find
\begin{eqnarray}
\langle\langle m_{U_{3}}^{2} \rangle\rangle & = & 0.132 ( 1-\frac{14.7}{{\cal{T'}}^{2}} ) m_{H}(0)^{2} + \left(\frac{\frac{1}{5}g_{1}^{2}-g_{2}^{2} +4|\lambda_{u_{3}}|^{2}}{\frac{3}{5}g_{1}^{2}+g_{2}^{2}}\right) \frac{\Delta^{2}m_{H}(0)^{2}}{{\cal{T}}^{2}}, \nonumber \\
\langle \langle m_{D_{3}}^{2} \rangle\rangle & = &  0.132 ( 1-\frac{14.7}{{\cal{T'}}^{2}} ) m_{H}(0)^{2}+\left(\frac{\frac{1}{5}g_{1}^{2}+g_{2}^{2}}{\frac{3}{5}g_{1}^{2}+g_{2}^{2}}\right) \frac{\Delta^{2}m_{H}(0)^{2}}{{\cal{T}}^{2}}, \nonumber \\
\langle\langle m_{U_{1,2}}^{2} \rangle\rangle & = & 0.465 ( 1-\frac{4.87}{{\cal{T'}}^{2}} ) m_{H}(0)^{2} + \left(\frac{\frac{1}{5}g_{1}^{2}-g_{2}^{2}}{\frac{3}{5}g_{1}^{2}+g_{2}^{2}}\right) \frac{\Delta^{2}m_{H}(0)^{2}}{{\cal{T}}^{2}}, \nonumber \\
\langle \langle m_{D_{1,2}}^{2} \rangle\rangle & = &  0.465 ( 1-\frac{4.87}{{\cal{T'}}^{2}} ) m_{H}(0)^{2} + \left(\frac{\frac{1}{5}g_{1}^{2}+g_{2}^{2}}{\frac{3}{5}g_{1}^{2}+g_{2}^{2}}\right) \frac{\Delta^{2}m_{H}(0)^{2}}{{\cal{T}}^{2}},
\label{211} \\
\langle \langle m_{u_{3}}^{2} \rangle \rangle &= & 0.374 ( 1-\frac{7.72}{{\cal{T'}}^{2}} )  m_{H}(0)^{2} -  \left(\frac{\frac{4}{5}g_{1}^{2}  -4|\lambda_{u_{3}}|^{2}}{\frac{3}{5}g_{1}^{2}+g_{2}^{2}}\right) \frac{\Delta^{2}m_{H}(0)^{2}}{{\cal{T}}^{2}}, \nonumber \\
\langle \langle m_{u_{1,2}}^{2} \rangle \rangle &= & 1.041 ( 1-\frac{3.42}{{\cal{T'}}^{2}} )   m_{H}(0)^{2} -  \left(\frac{\frac{4}{5}g_{1}^{2}}{\frac{3}{5}g_{1}^{2}+g_{2}^{2}}\right) \frac{\Delta^{2}m_{H}(0)^{2}}{{\cal{T}}^{2}}, \nonumber \\
\langle \langle m_{d_{i}}^{2} \rangle \rangle & = & 1.041 ( 1-\frac{3.42}{{\cal{T'}}^{2}} )  m_{H}(0)^{2}+  \left(\frac{\frac{2}{5}g_{1}^{2}}{\frac{3}{5}g_{1}^{2}+g_{2}^{2}}\right) \frac{\Delta^{2}m_{H}(0)^{2}}{{\cal{T}}^{2}} \nonumber \ .
\end{eqnarray}
Finally, using expression (\ref{180}) and taking $|\lambda_{u_{3}}|^{2}=1$, these squared masses become
\begin{eqnarray}
\langle\langle m_{U_{3}}^{2} \rangle\rangle & = & 0.132 m_{H}(0)^{2}(1-\frac{14.7-63.5 \Delta^{2}}{{\cal{T}}^{2}}) \simeq 0.132 m_{H}(0)^{2}, \nonumber \\
\langle \langle m_{D_{3}}^{2} \rangle\rangle & = & 0.132 m_{H}(0)^{2}(1-\frac{14.7-21.1 \Delta^{2}}{{\cal{T}}^{2}})  \simeq 0.132 m_{H}(0)^{2}, \nonumber \\
\langle\langle m_{U_{1,2}}^{2} \rangle\rangle & = & 0.465 m_{H}(0)^{2}(1-\frac{4.87-3.37 \Delta^{2}}{{\cal{T}}^{2}}) \simeq 0.465 m_{H}(0)^{2}, \nonumber \\
\langle \langle m_{D_{1,2}}^{2} \rangle\rangle & = & 0.465 m_{H}(0)^{2}(1-\frac{4.87-6.69 \Delta^{2}}{{\cal{T}}^{2}})  \simeq 0.465 m_{H}(0)^{2}, 
\label{212} \\
\langle \langle m_{u_{3}}^{2} \rangle \rangle &= & 0.374 m_{H}(0)^{2}(1-\frac{7.72-26.0 \Delta^{2}}{{\cal{T}}^{2}}) \simeq 0.374 m_{H}(0)^{2}, \nonumber \\
\langle \langle m_{u_{1,2}}^{2} \rangle \rangle &= & 1.041 m_{H}(0)^{2}(1-\frac{3.42-3.12 \Delta^{2}}{{\cal{T}}^{2}}) \simeq 1.041 m_{H}(0)^{2}, \nonumber \\
\langle \langle m_{d_{i}}^{2} \rangle \rangle & = & 1.041 m_{H}(0)^{2}(1-\frac{3.42-3.56 \Delta^{2}}{{\cal{T}}^{2}}) \simeq 1.041 m_{H}(0)^{2} \ \nonumber \ .
\end{eqnarray}
Note that the squark mass squares are  positive for for all ${\cal{T}}^{2} \stackrel{>}{\sim} 40$ and $0<\Delta^{2}<1$. The superpotential contributions to both third family up-squark masses can become significant when ${\cal{T}}^{2}$ and $\Delta^{2}$ are simultaneously in the lower and upper part of their range respectively. Be that as it may, for typical values of these parameters such contributions are negligible. Therefore, for simplicity, we ignore them when giving the leading order estimate of the squared masses in (\ref{212}). 

We conclude from (\ref{194}),(\ref{195}), (\ref{207}) and (\ref{212}) that 
\begin{eqnarray}
\langle \langle \nu_{1,2} \rangle \rangle  & = & 0 , \quad \langle \langle \nu_{3} \rangle \rangle = (1.05)\frac{2 m_{\nu}(0)}{\sqrt{\frac{3}{4}}g_{4}(t_{EW})},  \nonumber \\
\langle \langle e_{i} \rangle \rangle  & = & \langle \langle L_{i} \rangle \rangle = 0 \quad i=1,2,3
\label{213} \\
{\langle \langle} H'^{0} \rangle \rangle & = & \frac{2\Delta  m_{H}(0)}{{\cal{T}}\sqrt{\frac{3}{5}g_{1}(t_{EW})^{2}+g_{2}(t_{EW})^{2}}}, \nonumber \\
 \langle \langle H'^{+} \rangle \rangle & = & \langle \langle {\bar{H}}'^{0} \rangle \rangle  =  \langle \langle {\bar{H}}'^{-} \rangle \rangle = 0 
\nonumber
\end{eqnarray}
is a stable local minimum of the potential energy at scale $t_{EW}$.

\section{The $B$-$L$/Electroweak Hierarchy}

The vacuum state (\ref{213}) spontaneously breaks both $B$-$L$ and electroweak symmetry, and exhibits a distinct hierarchy between the two.  Color and electric charge are left unbroken. Using
(\ref{me5}), we see that the $B$-$L$/electroweak hierarchy, expressed as the ratio of the third right-handed sneutrino and Higgs vacuum expectation values, is
\begin{equation}
\frac{\langle \langle \nu_{3} \rangle \rangle}{\langle \langle H'^{0} \rangle \rangle} \simeq (0.976) \frac{\sqrt{\frac{3}{5}g_{1}^{2}+g_{2}^{2}}}{\sqrt{\frac{3}{4}}g_{4}} \frac{tan\beta}{\Delta} \ ,
\label{h1}
\end{equation}
where the gauge parameters are computed at $t_{EW}$. We have dropped the term proportional to $2.25(1-\Delta^{2})/tan\beta^{2}$ in (\ref{me5}) since it is always much less than unity in our parameter regime where  $tan\beta \stackrel{>}{\sim} 6.32$ and $0 < \Delta^{2} < 1$. 
 Note that for fixed $tan\beta$ the ratio of VEVs in (\ref{h1}) can be made arbitrarily large by fine-tuning $\Delta \rightarrow 0$. Conversely, by fine-tuning $\Delta \rightarrow 1$ this ratio approaches $2.22 \ tan\beta$, where we used (\ref{180}) and (\ref{g4tew}). A more natural value for $\Delta$ would lie in the middle of the range $0 \leq \Delta^{2} \leq 1$. For specificity, let us take $\Delta=\frac{1}{\sqrt{2}}$. In this case, the ratio (\ref{h1}) evaluated in the region $6.32 \leq tan\beta \leq 40$ is found to be
\begin{equation}
19.9 \leq \frac{\langle \langle \nu_{3} \rangle \rangle}{\langle \langle H'^{0} \rangle \rangle} \leq 
126 \ .
\label{h2}
\end{equation}

A second measure of the $B$-$L$/electroweak hierarchy is given by the ratio of the $B$-$L$ vector boson mass to the mass of the $Z$ boson. It follows from (\ref{101}),(\ref{me5}),(\ref{train1}) and(\ref{air2}) that
\begin{equation}
\frac{M_{A_{B-L}}}{M_{Z}}\simeq (1.95 )\frac{tan\beta}{\Delta} \ .
\label{h3}
\end{equation}
Note that if we take $\Delta \rightarrow 0$  this mass ratio becomes arbitrarily large, whereas if
$\Delta \rightarrow 1$, the upper bound in our approximation, then $\frac{M_{A_{B-L}}}{M_{Z}}$ is 
$1.95 \ tan\beta$. Again, using the more natural value $\Delta=\frac{1}{\sqrt{2}}$ and evaluating this mass ratio in the range $6.32 \leq tan\beta \leq 40$, one finds
\begin{equation}
17.5 \leq \frac{M_{A_{B-L}}}{M_{Z}} \leq 110 \ .
\label{h4}
\end{equation}

{\it We conclude from (\ref{h1}) and (\ref{h3}) that for typical values of $\Delta$, the vacuum (\ref{213}) exhibits a  $B$-$L$/electroweak hierarchy of ${\cal{O}}(10)$ to ${\cal{O}}(10^{2})$ over the physically interesting range $6.32 \leq tan\beta \leq 40$.} Let us review the reasons for the existence and magnitude of this hierarchy. First, initial conditions (\ref{39}),(\ref{40}),(\ref{80}),(\ref{81}) give emphasis to the right-handed sneutrinos by {\it not} requiring their masses be degenerate with the $L_{i}$ and $e_{i}$ soft masses. This enables the ${\cal{S}}'_{1}$ parameter (\ref{82}) not only to be non-vanishing but, in addition, to be large enough to dominate all contributions to the RGEs with the exception of the gluino mass terms. This drives $m_{\nu_{3}}^{2}$ negative and initiates $B$-$L$ breaking at scale $m_{\nu}$. Second, $B$ and $M_{3}$ (hence, $m_{H}(t_{EW})$) are chosen to satisfy constraints 
(\ref{buddy2}),(\ref{home1}) and (\ref{buddy3}) respectively, 
with $0<\Delta^{2}<1$. This insures electroweak breaking for positive $m_{H}^{2}$ at a scale proportional to $\Delta m_{H}(0)/ {\cal{T}}$.  The large value assumed for ${\cal{T}}$ implies that the non-vanishing VEV is largely in the $H^{0}$ direction, allowing one to identify ${\cal{T}}$, to leading order, with $\tan\beta$. Third, equation (\ref{me5}) insures that squark/slepton squared masses are positive at {\it all} scales. This guarantees that the electroweak breaking is substantially smaller than the $B$-$L$ scale, with the $B$-$L$/electroweak hierarchy proportional to $\tan\beta/\Delta$.  

To finish our analysis, we would like to emphasize that once the $\frac{M_{A_{B-L}}}{M_{Z}}$ 
ratio is fixed, that is, once the $B$-$L$ vector boson mass is measured, the spectrum of squarks and sleptons is completely determined. To see this, first recall that at $t_{EW}$
\begin{equation}  
M_{A_{B-L}} = \sqrt{2} g_{B-L}(t_{EW}) \ \langle \langle \nu_{3} \rangle \rangle \ .  
\label{yyy2}
\end{equation}
It then follows from (\ref{15}),(\ref{air2}) and (\ref{me5}) that 
\begin{equation}
m_{H}(0)=.362 \ M_{A_{B-L}} \ .
\label{h5}
\end{equation}
Here and henceforth we drop all $1/{\cal{T}'}^{2}$ terms since they are much less than unity in our parameter regime. It then follows from (\ref{207}) that the slepton masses, in the order of decreasing mass, are
\begin{eqnarray}
\langle \langle m_{\nu_{1,2}} \rangle\rangle &=& 5.26 \ M_{A_{B-L}} \ , \nonumber \\
\langle \langle m_{\nu_{3}} \rangle\rangle &=& 1.72 \ M_{A_{B-L}} \ , \label{h6} \\
\langle \langle m_{N_{i}} \rangle\rangle  =  \langle \langle m_{E_{i}} \rangle\rangle & = & 1.54 \ M_{A_{B-L}} \ , \nonumber \\
\langle \langle m_{e_{i}} \rangle\rangle &= & 1.30 \ M_{A_{B-L}} \ . \nonumber
\end{eqnarray}
Similarly, we see from (\ref{212}) that the squark masses, in descending order, are given by
\begin{eqnarray}
\langle \langle m_{u_{1,2}} \rangle\rangle  =  \langle \langle m_{d_{i}} \rangle\rangle & = & .614 \ M_{A_{B-L}} \ , \nonumber \\
\langle \langle m_{U_{1,2}} \rangle\rangle  =  \langle \langle m_{D_{1,2}} \rangle\rangle & = & .410 \ M_{A_{B-L}} \ , \label{h7} \\
\langle \langle m_{u_{3}} \rangle\rangle &=& .369 \ M_{A_{B-L}} \ , \nonumber \\
\langle \langle m_{U_{3}} \rangle\rangle  =  \langle \langle m_{D_{3}} \rangle\rangle & = & .219 \ M_{A_{B-L}} . \nonumber
\end{eqnarray}
Note that the  sleptons masses are on the order of $M_{A_{B-L}}$ and each is heavier than any squark. The squark masses are lighter, being on the order of about $20\%$-$60\%$ of
$M_{A_{B-L}}$, with the third family left-handed up and down squarks being the lightest.
We conclude that the radiative $B$-$L$/electroweak hierarchy also leads to a computable hierarchy in the squark/slepton masses.

\subsection*{Appendix A:  The Numerical Solution for $m_{H}(t)^{2}$}

In this Appendix, we present a numerical solution of equation (\ref{123A}).  This is accomplished using the numerical fitting and solving packages in Wolfram's Mathematica program \cite{Zf}.  The details of these packages are quite complex and we refer the reader to the documentation provided by Mathematica for further discussion.  

We begin by considering equation (\ref{123A}) with the property that $m_{H}^{2}$ is positive over the entire scaling range $t_{EW} \leq t \leq 0$. With this additional condition, (\ref{123A}) is equivalent to equation (\ref{129}). First note, using (\ref{24}), that (\ref{123A}) contains two arbitrary parameters,  $m_{H}(0)^{2}$ and $ |M_{3}(0)|^{2}$. The coefficient $m_{H}(0)^{2}$ is the value of the up-Higgs soft supersymmetry breaking mass at $t=0$ and, hence, is a natural input parameter. However, for the reasons discussed in the text, it is more transparent physically to input the value of $m_{H}^{2}$ at $t_{EW}$, given in (\ref{buddy2}),(\ref{home1}) and (\ref{buddy3}), rather than $|M_{3}(0)|^{2}$. To do this, one chooses ${\cal{T}}$ and $\Delta^{2}$ subject to the constraints ${\cal{T}}^{2} \apgt  40$ and $0<\Delta^{2}<1$. It then follows from (\ref{buddy2}) and (\ref{home1}) that we have completely specified the value of $m_{H}(t_{EW})^{2}$. For this choice of parameters, we solve the $m_{H}^{2}$ equation of motion (\ref{123A}) numerically, adjusting the value of $|M_{3}(0)|^{2}$ until the value of $m_{H}^{2}$ at $t_{EW}$ is given by $m_{H}(0)^{2}/{\cal{T}'}^{2}$. The plot of this value of $|M_{3}(0)|^{2}$ for a large number of choices of ${\cal{T}'}^{2}$ is shown in Figure 1. Fitting this result with a smooth curve, we find that 
\begin{equation}
|M_{3}(0)|^{2}=.0352(1-\frac{11.5}{{\cal{T}'}^{2}})m_{H}(0)^{2} \ .
\label{A1}
\end{equation}
This curve is also plotted in Figure 1 and closely reproduces the numerical data. This
justifies expression (\ref{cat1}) used in the text.

\begin{figure}
 \centering
 \includegraphics[scale=1]{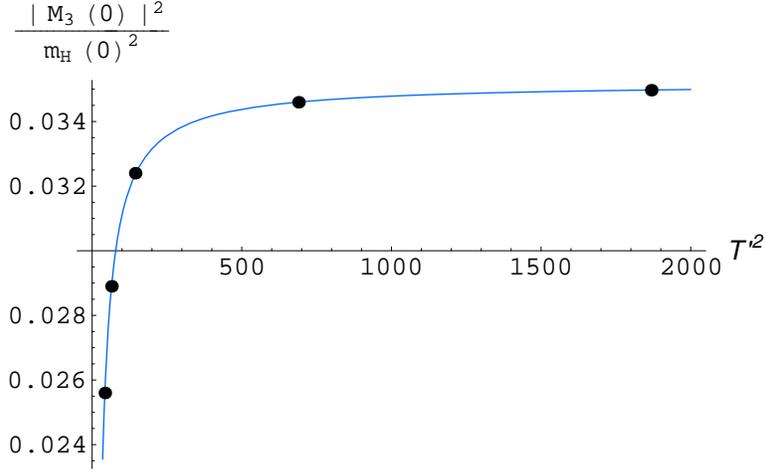}
 \caption{{\scriptsize  This plot shows a representative set of points of $|M_{3}(0)|^{2}/m_{H}(0)^{2}$ (black dots) for different values of ${\cal{T'}}^{2}$ as well as the accuracy of the fit of equation (\ref{A1}) (blue line) to  these representative points.  For simplicity we show only show a few points but many more were used in the generation of this fit.  The accuracy of this fit over the range of ${\cal{T'}}^{2}$ exceeds one percent at each data point (including those not shown).     }}
 \label{plot1}
\end{figure}

Having determined relation (\ref{A1}), one can now numerically solve the $m_{H}^{2}$ equation 
(\ref{123A}) for any choice of $m_{H}(0)^{2}$ and ${\cal{T}'}^{2}$. Numerical plots of $m_{H}^{2}/m_{H}(0)^{2}$  as a function of $t$ are shown in Figure 2 for several different choices of ${\cal{T}'}^{2}$. For each such graph, we fit the numerical results to a smooth curve. We find that for any allowed value of 
${\cal{T}'}^{2}$, the data is well represented by
\begin{equation}
m_{H}(t)^{2} = \big( 1 -(1 - \frac{1}{{\cal{T}'}^{2}} )( \frac{t_{EW}-b}{t_{EW}} ) ( \frac{t}{t-b} )\big)m_{H}(0)^{2}  \ ,
\label{A2}
\end{equation}
where b is a function of ${\cal{T'}}$ of the form
\begin{equation}
b({\cal{T'}}) =19.9(1-\frac{.186}{{\cal{T}'}-3.69}) \ .
\label{A3}
\end{equation}
Note that at $t=0$ and $t=t_{EW}$, $m_{H}^{2}$ is given by $m_{H}(0)^{2}$ and $m_{H}(0)^{2}/{\cal{T}'}^{2}$ respectively, as it must be. This smooth curve is plotted in Figure 2 for each of the choices of ${\cal{T}'}^{2}$and is seen to give a close fit to the numerical data. This justifies equations (\ref{funfit}),(\ref{funfit2}) used in the text.

\begin{figure}
 \centering
 \hspace*{-0.5in}
 \subfloat[${\cal{T'}} = 6.32$]{\label{plota}\includegraphics[scale=0.48]{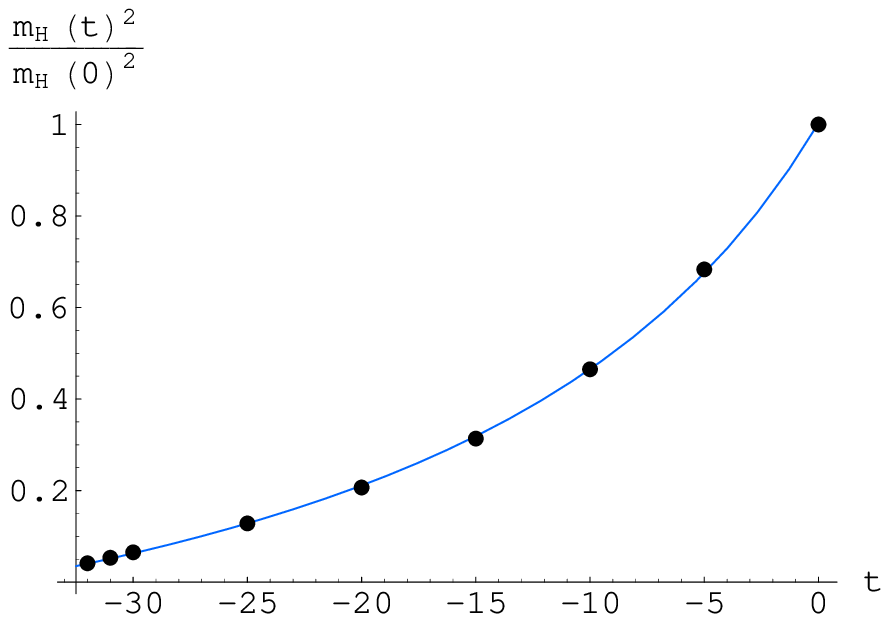}}
 \subfloat[${\cal{T'}} = 40$]{\label{plotb}\includegraphics[scale=0.48]{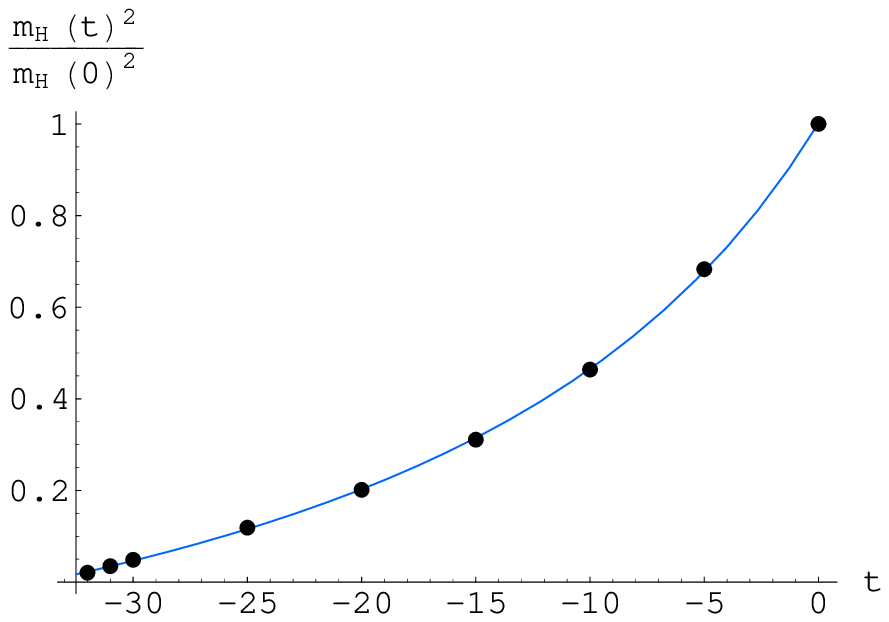}}
 \subfloat[${\cal{T'}} = 120$]{\label{plotc}\includegraphics[scale=0.48]{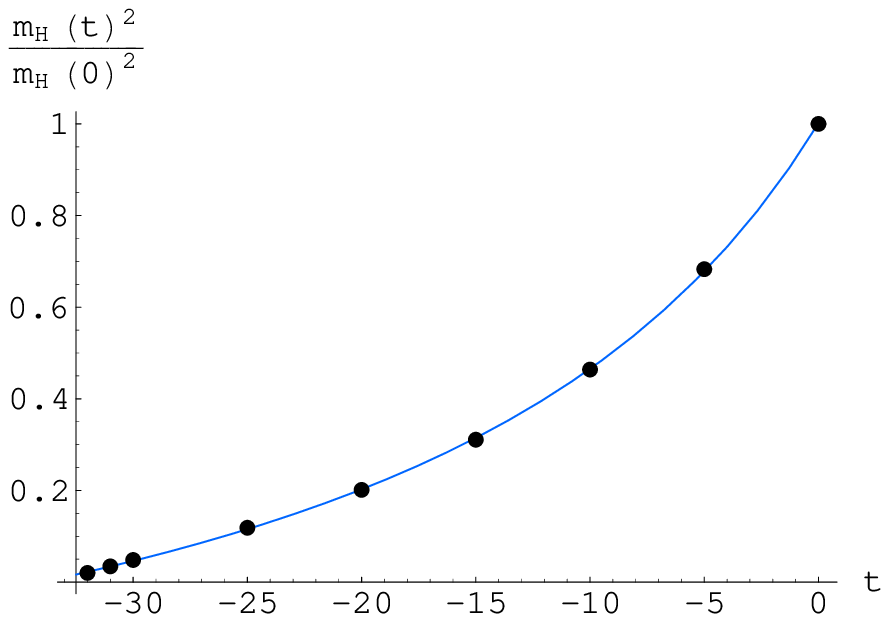}}
 \caption{{\scriptsize  These plots show representative points $m_{H}(t)^{2}/m_{H}(0)^{2}$ (black dots) obtained by numerical solution of (\ref{123A}) for different values of $t$ in our scaling range as well as the fit of equations (\ref{A2}),(\ref{A3}) (blue line) to these points.  We show this for three choices of ${\cal{T'}}$ spanning a wide range of physically interesting values. }}
 \label{plots2}
\end{figure}

\subsection*{Appendix B:  The Relationship Between $m_{\nu}(0)^{2}$ and  $m_{H}(0)^{2}$ }

As highlighted in the discussion following equations (\ref{me4}) and (\ref{206}), squark and slepton masses evaluated around the VEVs $\langle\langle \nu_{3} \rangle\rangle$, $\langle \langle H'^{0} \rangle \rangle$ generically depend on the two initial values, $m_{\nu}(0)^{2}$ and $m_{H}(0)^{2}$.  We want to explore the region of parameter space leading to positive squark/slepton masses  over the entire scaling range, thus simplifying the discussion of symmetry breaking.  This region of  parameter space can be specified in terms of a simple ${\cal{T}'}^{2}$ dependent relationship between these two initial conditions.  In this Appendix, we present a detailed derivation of this relationship.

Recall that the squark/slepton masses depend not only on the associated $m^{2}$ coefficients and their running, but on  contributions from the  $\langle\langle \nu_{3} \rangle\rangle$ and $\langle \langle H'^{0} \rangle \rangle$ VEVs as well.  These expectation values are only non-zero below certain scales, thus complicating the analysis.  Furthermore, both $m_{Q_{3}}^{2}$ and $m_{u_{3}}^{2}$ contain an $m_{H}^{2}$ term and, hence, require the numerical solution described in Appendix A.  As an explicit example of how the relationship between $m_{\nu}(0)^{2}$ and $m_{H}(0)^{2}$ effects the sign of the scalar mass terms, let us consider $\langle\langle m_{Q_{3}}^{2}\rangle\rangle$.  It follows from (\ref{143}), (\ref{118B}) and (\ref{209}) that
\begin{eqnarray}
\langle\langle m_{Q_{3}}^{2}\rangle\rangle & \simeq & \frac{1}{3}m_{H}^{2}-\frac{2}{3 {\pi}^{2}} \int_{0}^{t}{g_{3}^{2}|M_{3}|^{2}}+\frac{1}{64 {\pi}^{2}} \int_{0}^{t}g_{4}^{2}{\cal{S}'}_{1}   \nonumber  \\ 
&+&\frac{1}{6}m_{H}(0)^{2} + \frac{1}{4}g_{4}^{2}\langle\langle \nu_{3} \rangle\rangle^{2} \label{abc} 
+ \frac{1}{4} ( \frac{1}{5} g_{1}^{2} \mp  g_{2}^{2} )  \langle\langle H'^{0} \rangle\rangle ^{2} \ , 
\end{eqnarray}
where $\mp$ indicates that for the up field $U_{3}$ and the down field $D_{3}$ of the doublet $Q_{3}$ one  uses a minus sign and plus sign respectively.  
\begin{figure}
 \centering
 \hspace*{-0.12 in}
 \subfloat[]{\label{plot1}\includegraphics[scale=0.65]{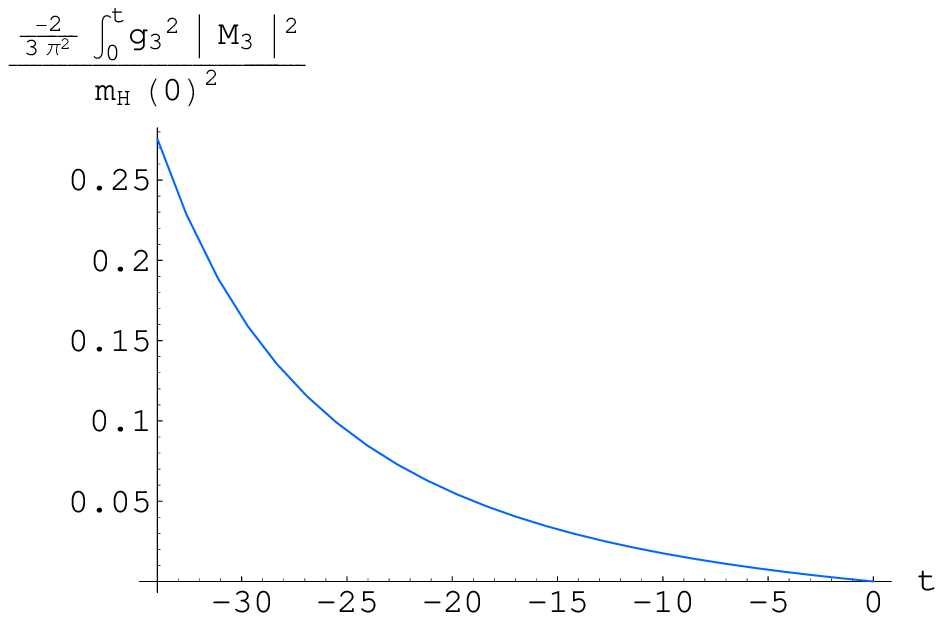}}
 \subfloat[]{\label{plot2}\includegraphics[scale=0.65]{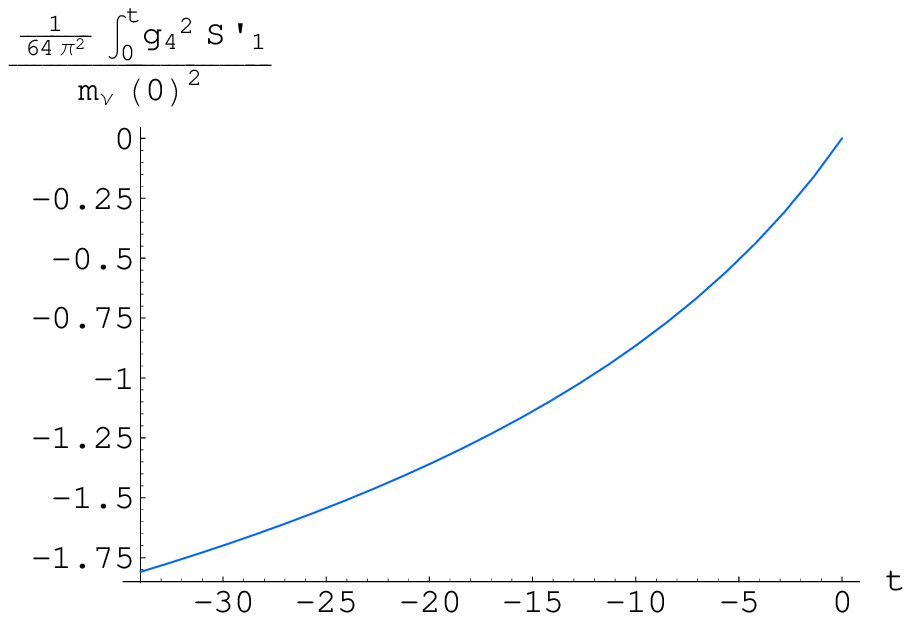}}
 \caption{{\scriptsize Graph (a) is a plot of the term $-\frac{2}{3 {\pi}^{2}} \int_{0}^{t}{g_{3}^{2}|M_{3}|^{2}}$  over the entire scaling range taking ${\cal{T'}}^{2}$=40. Graph  (b) is the term $\frac{1}{64 {\pi}^{2}} \int_{0}^{t}g_{4}^{2}{\cal{S}'}_{1}$ plotted over the same range. }}
 \label{plots1}
\end{figure}
Let us discuss each term in this equation and its value over the entire scaling range.  The first term is  $m_{H}^{2}$ derived in Appendix A and given in (\ref{A2}), (\ref{A3}).  It is proportional to $m_{H}(0)^{2}$ and depends explicitly on the value of ${\cal{T}'}$. We plotted it over the scaling range for three different values of ${\cal{T}'}$ in that Appendix.  Next we have the integrals containing $|M_{3}|^{2}$ and $S'_{1}$.  These can be evaluated for arbitrary $t$ using equations (\ref{120}),(\ref{cat1}) and (\ref{93}),(\ref{121}) respectively. Note that the first is ${\cal{T}'}^{2}$  and $m_{H}(0)^{2}$ dependent, while the second is $m_{\nu}(0)^{2}$ dependent. Figure \ref{plots1} plots their values over the scaling range, where in graph (a) we have taken ${\cal{T}'}=40$ for specificity.  Next we have the sneutrino VEV term.  It is given for any scale $t$ by minimizing the purely $\nu_{3}$ part of potential (\ref{94}) using equations 
(\ref{19}),(\ref{55}) and (\ref{93}) respectively and is plotted in Figure \ref{plot2} over the scaling range.  Note that this term depend on $m_{\nu}(0)^{2}$.  Lastly, we have the Higgs VEV term. This turns out to give a very small contribution, so we omit it henceforth.

\begin{figure}
 \centering
 \includegraphics[scale=0.8]{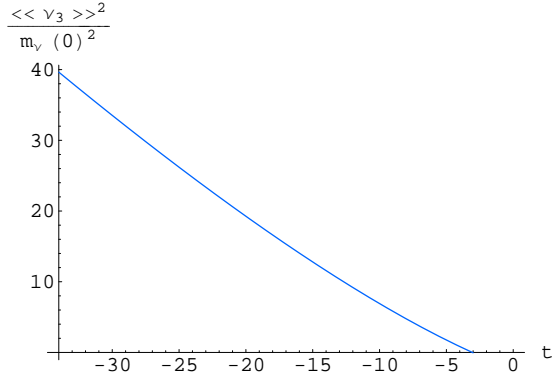}
 \caption{{\scriptsize  In this graph, $\langle\langle\nu_{3}\rangle\rangle^{2}$/$m_{\nu}(0)^{2}$ is plotted over the entire scaling range.   }}
 \label{plot2}
\end{figure}

\begin{figure}
 \centering
 \includegraphics[scale=0.8]{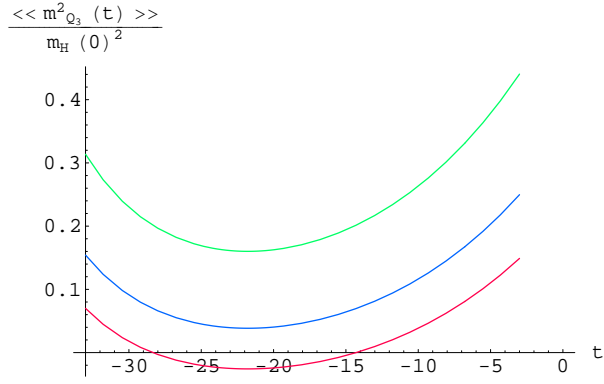}
 \caption{{\scriptsize  In this graph, we demonstrate the dependence of the running of $\langle \langle m_{Q_{3}}^{2} \rangle \rangle$ on $D^{2}$.  We plot $\langle \langle m_{Q_{3}}^{2} \rangle \rangle$/$m_{H}(0)^{2}$ for D=0.75 (green line), D=0.85 (blue line), and D=0.95 (red line).  It is apparent from this plot that as $D$ increases, $\langle \langle m_{Q_{3}}^{2} \rangle \rangle$ goes negative.  For this plot, we took ${\cal{T}'}^{2} = 40$.     }}
 \label{plot6}
\end{figure}

\begin{figure}
 \centering
 \includegraphics[scale=0.8]{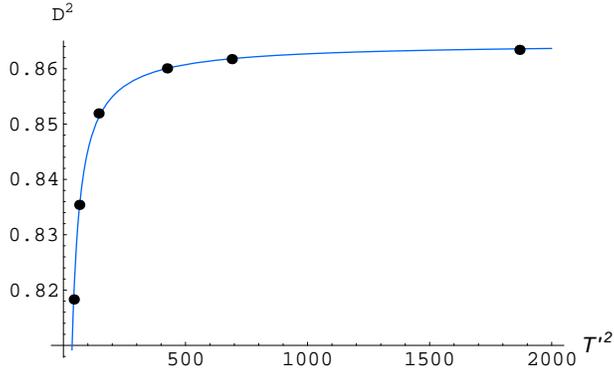}
 \caption{{\scriptsize  This plot shows representative points of $D^{2}$ (black dots) for different values of ${\cal{T'}}^{2}$ and the fit of equation (\ref{ias1}) (blue line) to  these points.  The accuracy of this fit over the range of ${\cal{T'}}^{2}$ exceeds one percent at each data point.   }}
 \label{plot5}
\end{figure}

Note that each of the above terms is proportional  to either $m_{\nu}(0)^{2}$ or $m_{H}(0)^{2}$.  Furthermore, three of the terms are everywhere positive whereas the two remaining terms, which are proportional to $m_{\nu}(0)^{2}$, are everywhere negative.  It is easy to demonstrate that for a sufficiently large ratio of  $m_{\nu}(0)^{2}$ to  $m_{H}(0)^{2}$, $\langle \langle m_{Q_{3}}^{2} \rangle \rangle$ could be negative somewhere in the scaling range.  To investigate this, we first define
\begin{equation}
 m_{\nu}(0)^{2} = D^{2} m_{H}(0)^{2} \ .
\label{BD} 
\end{equation}
This allows us to write each term as a function of one of these initial masses, which we choose here to be $m_{\nu}(0)^{2}$, and their ratio $D^{2}$. Inserting the renormalization group expression for the scaling of each term into (\ref{abc}), we find that 
\begin{align}
\langle \langle m_{Q_{3}}^{2} \rangle \rangle  \simeq & \Bigg\lbrace\frac{1}{3}\big( 1 -(1 - \frac{1}{{\cal{T}'}^{2}} )( \frac{t_{EW}-b}{t_{EW}} ) ( \frac{t}{t-b} )\big) \frac{m_{\nu}(0)^{2}}{D^{2}} \Bigg\rbrace \nonumber \\ 
 -&\Bigg\lbrace\frac{8}{3b_{3}}( \frac{1}{(1-\frac{g(0)^{2}b_{3}t}{8 {\pi}^{2}})^{2}} -1)  0.0352(1-\frac{11.5)}{{\cal{T'}}^{2}})\frac{m_{\nu}(0)^{2}}{ D^{2}}\Bigg\rbrace \nonumber \\  
 +&\Bigg\lbrace\frac{1}{18}( \frac{1}{(1-\frac{g(0)^{2}b_{4}t}{8 {\pi}^{2}})^{\frac{9}{4b_{4}}}}-1)149 \ m_{\nu}(0)^{2}\Bigg\rbrace +\frac{m_{\nu}(0)^{2}}{6  D^{2}}  \nonumber  \\
- &\Bigg\lbrace m_{\nu}(0)^{2} - \frac{1}{6} ( 1 - ( 1- \frac{g(0)^{2} b_{4} t  }{ 8 \pi^{2} } )^{-\frac{9}{4 b_{4}}} )  149 m_{\nu}(0)^{2}\\ 
+ &\sqrt{\frac{3}{4}}  \frac{g(0)}{( 1-\frac{g(0)^{2} b_{4} t}{8 \pi^{2}} )^{\frac{1}{2}}} \bigg( \frac{-2}{g(0) (b_{4}-\frac{9}{2})}\nonumber  \\ 
\times & \bigg( \Big( 1 - \frac{g(0)^{2} b_{4} t}{8 \pi^{2}} \Big)^{\frac{1}{2} ( 1 - \frac{9}{2b_{4}})} -1 \bigg) \bigg)  149 m_{\nu}(0)^{2} \Bigg\rbrace   \nonumber \ .
\end{align}
We have preserved the original ordering of terms for ease of reference.  Recall that the 
coefficient $b$ is defined in (\ref{A3}).  Note that we can factor out $m_{\nu}(0)^{2}$ and, hence, the properties of this equation are controlled by the value of $D^{2}$ and ${\cal{T}'}^{2}$. Choose a fixed value for ${\cal{T}'}^{2}$. If we initially assume that $D^{2} \ll 1$, that is, $m_{\nu}(0)^{2} \ll m_{H}(0)^{2}$, then $\langle \langle m_{Q_{3}}^{2} \rangle \rangle$ is everywhere positive. However, as we increase $D^{2}$ the value of  $\langle \langle m_{Q_{3}}^{2} \rangle \rangle$ becomes smaller, eventually touching zero at some point $t$ in the scaling regime. For still larger $D^{2}$ the value of $\langle \langle m_{Q_{3}}^{2} \rangle \rangle$ at this point, and in a range around it, is negative, see Figure \ref{plot6}.
Record the value of $D^{2}$ at which $\langle \langle m_{Q_{3}}^{2} \rangle \rangle$ just touches zero and repeat this for a set of choices of ${\cal{T}'}^{2}$. This is plotted in Figure \ref{plot5}. We find that a smooth curve fitting this data is given by
\begin{equation}
D^{2}=0.864(1- \frac{2.25}{{\cal{T}'}^{2}} )
\label{ias1}
\end{equation}
or, equivalently, using (\ref{BD}) that
\begin{equation}
m_{\nu}(0)^{2}=0.864(1- \frac{2.25}{{\cal{T}'}^{2}} )m_{H}(0)^{2} \ .
\label{ias2}
\end{equation}

This equation specifies the largest value of $m_{\nu}(0)^{2}$ relative to $m_{H}(0)^{2}$ for which $\langle \langle m_{Q_{3}}^{2} \rangle \rangle$ is everywhere non-negative.
{\it An investigation of all other other squark and slepton mass squares evaluated around the $B$-$L$ and Higgs VEVs shows that they are everywhere positive whenever $\langle\langle m_{Q_{3}}^{2} \rangle\rangle$ is everywhere non-negative.}  That is, for $m_{\nu}(0)^{2}$ and $m_{H}(0)^{2}$ satisfying relation (\ref{ias2}), all squark/slepton masses are everywhere positive, as desired. Furthermore, this is the largest value of $m_{\nu}(0)^{2}$ relative to $m_{H}(0)^{2}$ for which this is the case. This is chosen so that the $B$-$L$/EW hierarchy, which is proportional to $m_{\nu}(0)/m_{H}(0)$, is as large as possible within this context. This justifies our use of equation (\ref{me5}) in the text.

\subsection*{Appendix C: Comparison to the Standard Formalism and the Higgsino Mass}

In this Appendix, we compare our analysis of the Higgs section with the standard MSSM results given, for example, in ``A Supersymmetric Primer'' by Stephen Martin \cite{Zc}. Restoring the $\mu$ paramter, that is, not assuming it is necessarily sub-leading, the complete potential for the neutral Higgs fields becomes 
\begin{eqnarray}
 V &=& ( |\mu |^{2} +  m_{H}^{2} ) |H^{0}|^{2}+ ( |\mu |^{2} + m_{\bar{H}}^{2} ) |{\bar{H}}^{0}|^{2} -  B(H^{0}{\bar{H}}^{0}+hc) \nonumber  \\
& + &\frac{1}{8}(\frac{3}{5}g_{1}^{2} + g_{2}^{2} )  (|H^{0}|^{2} - |{\bar{H}}^{0}|^{2})^{2} 
\label{c1}.
\end{eqnarray}  
\noindent Note that this potential is written in terms of the original Higgs fields, not the mass eigenstates at the origin we used in our analysis. The two constraints that must be satisfied to have a stable, non-vanishing vacuum of (\ref{c1}) are presented as equations (7.3) and (7.4) in the Supersymmetric Primer.
The two minimization equations are given in (7.10) and (7.11) of that review. Note that each of these equations should be evaluated at the electroweak scale $t_{EW}$. We now show, to the order in ${\cal{T}}^{-1}$ we are working to, that our vacuum satisfies all four of these equations.\\
\indent We begin by examining the constraint (7.3) in~\cite{Zc}. This equation, which ensures that the potential is bounded from below, and is given by
\begin{equation}
2B < 2 |\mu |^{2} + m_{H}^{2} + m_{\bar{H}}^{2} \ .
\label{c2}
\end{equation}
Using (\ref{buddy3}), this can be rewritten in the form
\begin{equation}
  \frac{1}{{\cal{T}}} = |\mu |^{2} + \frac{m_{H}^{2}}{m_{\bar{H}}^{2} -m_{H}^{2}} + \frac{1}{2} \ .
\label{c3}
\end{equation}
It then follows from (\ref{150}),(\ref{buddy2}),(\ref{home1}) and (\ref{buddy3}) that this inequality is satisfied by our vacuum over the entire range ${\cal{T}} \stackrel{>}{\sim} 6.32$ and $0 < \Delta^{2} < 1$ 
for {\it any value of the parameter $\mu$.}
Next, consider the constraint equation (7.4) in \cite{Zc} given by
\begin{equation}
B^{2} < (|\mu |^{2} +  m_{H}^{2} ) ( |\mu |^{2} + m_{\bar{H}}^{2} ) \ .
\label{c5}
\end{equation}
This ensures that the origin will not be a stable minimum of the potential.  
Using (\ref{150}),(\ref{buddy2}),(\ref{home1}) and (\ref{buddy3}) this becomes
\begin{equation}
 1 > ( 1 + \frac { |\mu |^{2} }{  m_{H}^{2} } ) ( 1 + \frac{ |\mu |^{2} }{ m_{H}^{2}(0)} ) \frac{ 1- \Delta^{2}  }{ ( 1- \frac{1-\Delta^{2}}{{\cal{T}}^{2}} )^{2}} \ .
\label{c6}
\end{equation}
For the time being, {\it take $\mu$ to be sub-dominant to the Higgs mass parameters} and ignore it, as we did in the text.  Later in this Appendix we will keep it in our analysis and observe the consequences.  
We are left with a simple inequality 
\begin{equation}
( 1- \frac{1-\Delta^{2}}{{\cal{T}}^{2}} )^{2} > 1- \Delta^{2} \ .
\label{c6a}
\end{equation}
We want to examine if this holds over the entire range $0 < \Delta^{2} < 1$ and $ {\cal{T}} \stackrel{>}{\sim} 6.32$ of our vacuum solution.  Note that (\ref{c6a}) is trivially satisfied  as $\Delta^{2} \rightarrow 1$.  However, in the $\Delta^{2} \rightarrow 0$ limit this inequality clearly does {\it not} hold.  To find what bound is set by the above equation, solve (\ref{c6a}) for $\Delta^{2}$ keeping the relevant root. We find that the inequality will be satisfied for
\begin{equation}
 \Delta^{2} > 0 + \frac{2}{{\cal{T}}^{2}}  +{\cal{O}}(\frac{1}{{\cal{T}}^{4}})
\label{c7}
\end{equation}
for any value of ${\cal{T}}$.
To the order we are working in this paper, $0$ is in fact the lower bound and inequality (\ref{c6a}) is satisfied over the whole range of $\Delta^{2}$.  However, to next order in ${\cal{T}}^{-1}$, a constraint appears.  This is due to the fact that we only took the first order approximation in equation (\ref{149}).  Had we kept higher orders, then relation (\ref{home2}) would be more complicated and, hence, the lower bound on $\Delta^{2}$ different.  It is straightforward to verify that, to higher order 
in ${\cal{T}}^{-1}$, the lower bound would be identical to (\ref{c7}). We conclude that, to the order we are working,  constraint (7.4) in \cite{Zc} is satisfied over the entire range ${\cal{T}} \stackrel{>}{\sim} 6.32$ and $0<\Delta^{2}<1$ of our vacuum.

We now show that identities (7.10) and (7.11) in \cite{Zc}, which are derived from the minimization conditions of potential (\ref{c1}), are also satisfied by our vacuum.  First consider (7.10) given by
\begin{equation}
  sin2 \beta = \frac{2 B}{ m_{H}^{2} + m_{\bar{H}}^{2} +2 |\mu |^{2}}  \ .
\label{c8}
\end{equation}
This equation can be rewritten as
\begin{equation}
  tan \beta= \frac{2}{{\cal{C}}(1-(1-\frac{4}{{\cal{C}}^{2}})^{1/2})} \ ,
\label{c9}
\end{equation}
where 
\begin{equation}
{\cal{C}}^{-1}=\left(\frac{B}{m_{H}^{2} + m_{\bar{H}}^{2} +2 |\mu |^{2}}\right) .
\label{c9a}
\end{equation}
Expressing ${\cal{C}}$ in terms of ${\cal{T}}$ using (\ref{buddy3}), we find that (\ref{c9}) becomes
\begin{equation}
 tan\beta=  {\cal{T}}(1+\frac{2}{{\cal{T}}^{2}}+ {\cal{O}}(\frac{1}{{\cal{T}}^{4}})) \ .
\label{c10}
\end{equation}
Hence, to leading order, $ {\cal{T}} = tan \beta$. This is identical to expression (\ref{199}) in our vacuum solution. Note that this is true {\it for any value of the parameter $\mu$.} It also follows from (\ref{c10}) that  
to higher order in ${\cal{T}}^{-1}$, the relationship of ${\cal{T}}$ to $tan\beta$ becomes non-linear.
Finally, consider the minimization equation (7.11) in \cite{Zc}. For our purposes, this is most conveniently re-expressed for large $tan\beta$ in (7.12) as
\begin{equation}
m_{Z}^{2}=-2(m_{H}^{2}+|\mu|^{2})+\frac{2}{tan^{2}\beta}(m_{\bar{H}}^{2}-m_{H}^{2})+{\cal{O}}(\frac{1}{tan^{4}\beta}) \ .
\label{c10a}
\end{equation}
It is clear from (\ref{150}),(\ref{buddy2}),(\ref{train1}),(\ref{199}) that as long as {\it $\mu$ is chosen to be sub-dominant to $m_{H}$}, our vacuum explicity solves (\ref{c10a}). We conclude that, to the order we are working,  minimization equations (7.10) and (7.11) in \cite{Zc} are satisfied over the entire range ${\cal{T}} \stackrel{>}{\sim} 6.32$ and $0<\Delta^{2}<1$ of our vacuum.

An assumption in this paper was that $\mu$, while non-vanishing, was sub-leading to the Higgs mass parameters at $t_{EW}$ and, hence, ignorable in the vacuum solution to leading order.  Can one quantify how much this  assumption restricts the size of $\mu$ and, hence, physical quantities such as the Higgsino mass? 
To do this, note from this assumption and (\ref{150}),(\ref{buddy2}) and (\ref{home1}) that
\begin{equation}
|\mu|^{2} \ll m_{H}^{2}\simeq \frac{1-\Delta^{2}}{{\cal{T}}^{2}}m_{H}(0)^{2} \ll m_{\bar{H}}^{2}\simeq m_{H}(0)
\label{c10b}
\end{equation}
in our vacuum. Therefore, a reasonable parameterization of $\mu$ which is sub-leading to the Higgs masses is
\begin{equation}
 \frac{|\mu |^{2}}{m_{H}^{2}} =  \frac{\alpha (1-\Delta^{2})}{{\cal{T}}^{2}} \ ,
\label{c11}
\end{equation}
where $\alpha$ is a constant of ${\cal{O}}(1)$.  It then follows from (\ref{buddy2}) that
\begin{equation}
 \frac{|\mu |^{2}}{m_{H}^{2}(0)} =  \frac{\alpha (1-\Delta^{2})^{2}}{{\cal{T}}^{4}} \ .
\label{c12}
\end{equation}
Putting this into inequality  (\ref{c6}), one finds the condition
\begin{equation}
 \Delta^{2} > \frac{2 + \alpha}{{\cal{T}}^{2}} 
\label{c13}
\end{equation}
to the first non-vanishing order in ${\cal{T}}^{-1}$.
To see how this limits the size of $\mu$, calculate the ratio of $\mu$ to $M_{Z}$ using (\ref{train1}) and (\ref{c11}).  We find
\begin{equation}
 \frac{|\mu |}{M_{Z}} = \sqrt{ \frac{\alpha}{2}} \frac{(1-\Delta^{2})}{\Delta}  \frac{1}{{\cal{T}}} \ .
\end{equation}
Applying constraint (\ref{c13}), it follows that
\begin{equation}
\frac{|\mu |}{M_{Z}} < \sqrt{ \frac{\alpha}{2(\alpha + 2)}} ( 1 - \frac{2+\alpha}{{\cal{T}}^{2}} )
\end{equation}
which, for the range  ${\cal{T}} \stackrel{>}{\sim} 6.32$ and taking, for example, $\alpha = 1 $, gives 
\begin{equation}
|\mu | < \frac{1}{\sqrt{6}} M_{Z} = 37.5 GeV \ .   
\end{equation}
We conclude that, even subject to our assumption that it be sub-leading to the Higgs mass parameters at $t_{EW}$, the $\mu$ parameter and, hence, such quantities as the Higgsino mass, can be relatively large, with an upper bound approaching 40\% of the $Z$-mass.


\section*{Acknowledgments}
B.A.O. would like to thank Nima Arkani-Hamed and Gil Paz for helpful discussions. B.A.O. is grateful to the Institute for Advanced Study and the Ambrose Monell Foundation for support.
M.A. would like to thank the Institute for Advanced Study for its hospitality.
The work of M.A. and B.A.O. is supported in part by the DOE under contract No. DE-AC02-76-ER-03071.
B.A.O. acknowledges partial support from the NSF RTG grant DMS-0636606.

\end{document}